\documentclass{pasj00}
\usepackage{multirow}
\usepackage[varg]{txfonts}
\draft

\begin{document}
\SetRunningHead{Mitsuishi et al.}{
An X-ray study of hot gas in NGC 253}
\Received{2012/07/23}
\Accepted{2012/12/03}

\title{An X-ray study of the galactic-scale starburst-driven outflow in NGC 253
}

\author{Ikuyuki  \textsc{Mitsuishi}$^{1, 2}$, Noriko Y. Yamasaki$^2$, 
Yoh Takei$^2$}

\affil{$^1$ Department of Physics, Tokyo Metropolitan University,
1-1 Minami-Osawa Hachioji Tokyo 192-0397 JAPAN\\
$^2$ Institute of Space and Astronautical Science, Japan Aerospace Exploration Agency (ISAS/JAXA),\\
3-1-1 Yoshinodai, Chuo-ku Sagamihara, Kanagawa, 252-5210}
\email{mitsuisi@phys.se.tmu.ac.jp}





%


%


\maketitle

\begin{abstract}
X-ray properties of hot interstellar gas in a bright, nearby edge-on starburst galaxy NGC 253 
were investigated to gain a further understanding of starburst-driven outflow activity 
by {\it XMM-Newton} and {\it Suzaku}.
Spectroscopic analysis for three regions of the galaxy characterized by multiwavelength observations 
i.e., the superwind region, the disk region and the halo region, was conducted. 
Various emission lines from O, Ne, Mg, Si and Fe were observed in the spectra of each region. 
The hot gas was represented by two thin thermal plasmas 
with temperatures of $kT$ $\sim$0.2 and $\sim$0.6 keV.
Abundance ratios i.e., O/Fe, Ne/Fe, Mg/Fe and Si/Fe, are consistent 
between three regions, which suggests the common origin of the hot gas.
The abundance patterns are consistent with those of type II supernova ejecta, 
indicating that the starburst activity in the central region 
provides metals toward the halo through a galactic-scale starburst-driven outflow.
The energetics also can support this indication 
on condition that 0.01--50 $\eta^{1/2}$ \% of the total emission in the nuclear region has flowed to the halo region.
To constrain the dynamics of hot interstellar gas, 
surface brightness and hardness ratio profiles 
which trace the density and temperature  were extracted.
Assuming a simple polytropic equation of state of gas, $T$$\rho^{1-\gamma}$ = const,  
we constrained the physical condition.
$\gamma$ is consistent with 5/3 at the hot disk of $<$3 kpc from the center along with the minor axis  
and $T$ is constant ($\gamma$ = 1) in the halo whose distance is between 3 and 10 kpc 
from the center.
It is suggested that the hot gas expands adiabatically from the central region towards the halo region 
while it moves as free expansion from the inner part of the halo towards the outer part of the halo as the outflow.
We constrained the outflow velocity to be $>$100 km s$^{-1}$ 
from the observed temperature gradient in the halo.
In comparison with the escape velocity of $\sim$220 km s$^{-1}$ for NGC 253, 
it is indicated that the hot interstellar gas can escape from the gravitational potential of NGC 253 
by combining the outflow velocity and the thermal velocity.
\end{abstract}

\section{Introduction}
\label{SEC:introduction}
It has long been known that the intergalactic medium contains substantial amount of metals 
(e.g., \citet{1996AJ....112..335S,2001ApJ...560..599A}).
The metals are thought to originate from stars in galaxies.
However, the mechanism of how the metals are expelled 
from the galaxies to an intergalactic space is still not clear and arguable.
Galactic-scale outflows have been regarded to play an important role in 
transporting materials into the intergalactic space.
As a power source of galactic-scale outflows, 
starburst activity is one of the most plausible candidates 
(e.g., \citet{2002ASPC..253..387S,2003RMxAC..17...47H,2005ARA&A..43..769V}).
Some galaxies have been experiencing starbursts (e.g., \citet{1998ApJ...498..541K,2004ApJS..151..193S}) and 
starburst activity itself is considered to be common and universal in a life of a galaxy 
(e.g., \citet{2001ApJ...554..981P,2003ApJ...588...65S,2010ApJ...724...49M}).
According to a scenario of a galactic-scale starburst-driven outflow, 
thermal pressure builds up in an interstellar medium (ISM) of a galaxy 
due to a supernova shock and then removes a fraction of the ISM \citep{2005ARA&A..43..769V}.
Finally, hot gas expands and breaks out of the disk and the halo of the host galaxy as the outflow, 
if the pressure is large enough to 
escape the gravitational potential of the disk and the host galaxy.
This heated outflowing hot gas emits mainly X-rays and 
therefore X-rays can be a tracer of outflowing hot gas itself.
To test this scenario observationally, 
chemical abundance is a good probe  
because great deal of type II supernovae produced by massive stars 
associated with starburst activity provide abundant $\alpha$ elements selectively.
Recently, 
abundance patterns for several starburst galaxies, i.e., NGC 4631, M82 and NGC 3079, were reported 
\citep{ngc4631-yamasaki,2011PASJ...63S.913K,2012arXiv1205.6005K}.
$\alpha$ elements such as O, Ne, Mg, Si and S are observed abundantly 
even outside the disk and 
they argue that the observed ISM outside the disk originates from the inner starburst region 
contaminated by type II supernovae.
However, abundance patterns were not accurately constrained in some of previously 
studied galaxies due to poor statistics of NGC 4631 and 
NGC 3079 except for a central region and contamination of charge-exchange emission in M82 disk.
To pursue the origin of hot gas in the halo chemically, 
a continuous measurement of abundance patterns from the inner starburst region 
to the halo is important.

NGC 253 is a bright nearby well-studied starburst galaxy.
The edge-on orientation and 
its large apparent diameter make it a suitable target to study the spatial distribution 
of emission features and the outflow from galaxies.
Actually, the 100 pc-scale outflow of ionized gas nearby the nuclear region with velocity of 100--300 km s$^{-1}$ 
is reported by an H$\alpha$ observation \citep{2011MNRAS.414.3719W}.
%
%
This H$\alpha$ outflow is very similar in morphology to the known X-ray emitting hot gas 
on scales down to $\leq$20 pc \citep{2000AJ....120.2965S}.
$AKARI$ revealed that NGC 253 holds a uniquely-shaped far-infrared halo reaching out to $\sim$9 kpc 
perpendicular to the disk region \citep{2009ApJ...698L.125K}.
Interestingly, the FIR halo clearly traces an X-ray emitting gas \citep{ngc253-xmm-halo}, 
which suggests that dust in the halo comes from the inner region together with X-ray emitting hot gas.
In \citet{2009ApJ...698L.125K}, they estimate an averaged dust outflow velocity to be 300-2000 km s$^{-1}$ and predict that 
dust would be escaping from the gravitational potential of NGC 253.
%
%

\citet{2011ApJ...742L..31M} studied intensively the central disk of NGC 253 
with {\it Chandra}, {\it XMM-Newton} and {\it Suzaku}.
The hard X-ray emitting gas associated with the densest molecular clouds \citep{ngc253-radio-sakamoto} 
are suggested to originate from the starburst activity.
This central hot gas can be a source of hot gas observed outside the nuclear region 
through the outflow.
In this paper, we analyze X-ray emission from whole NGC 253 with {\it XMM-Newton} and {\it Suzaku}, 
to verify the galactic-scale starburst-driven outflow chemically.
We extract abundance patterns for three characteristic regions of NGC 253 and 
discuss the origin of hot gas in the halo region.
Then, we also constrain 
physical conditions of hot gas in the disk and the halo regions 
and finally discuss the possibility of the outflow toward the intergalactic space.

Throughout this paper, we adopt the distance of 3.4 Mpc \citep{ngc253-distance} to NGC 253 
which corresponds to 16 pc arcsec$^{-1}$ at the redshift of 8.1$\times$10$^{-4}$. 
Unless otherwise specified, all errors are at 90 $\%$ confidence level in text and tables, 
and at 1 $\sigma$ in figures.
   

\section{Observations \& Data Reduction}
\label{SEC:obs-data-reduction}

To pursue the origin of hot gas in the halo region, 
we studied three regions in NGC 253 characterized by multiple wavelength observations 
as shown in Figure \ref{fig:three-regions}.
The first one is a region (hereafter the $superwind$ region) 
where an outflowing ionized gas is detected in the H$\alpha$ observation \citep{2011MNRAS.414.3719W} 
with velocity of 100--300 km s$^{-1}$ from the central region toward an outside  
perpendicular to the disk of the order of several 100 pc.
This outflow is considered to originate from the nuclear region.
Thus, we extracted the 1.4$\times$0.9 kpc$^2$ area including this outflow region.
The second one is a region (the $disk$ region) characterized by 
optical B band which is mainly attributed to massive stars 
associated with the starburst activity.
The optical disk region is elliptically distributed with major/minor axes of 14$\times$3 kpc$^2$ 
\citep{1980ApJ...239...54P}.
We adopted this optical disk region as the $disk$ region.
The last one is a region (the $halo$ region) characterized by 
a diffuse FIR emission extending up to 9 kpc perpendicular to the disk 
\citep{2009ApJ...698L.125K}.
From its spatial coincidence with an X-ray emitting plasma, 
the extended FIR emission is considered to be transported from the central starburst region 
into the halo by the outflow associated with the starburst activity.
The $halo$ region was outside the $disk$ region (15$\times$7.4 kpc$^2$).

For spectral analysis of the $superwind$ region and the $disk$ region, 
we utilized {\it XMM-Newton}.
{\it XMM-Newton} has the angular resolution of $\sim$10$''$ 
corresponding to $\sim$160 pc, which is beneficial in removing a contamination 
from point sources and extracting abundance patterns accurately 
with a good sensitivity for soft emission lines.
For spectral analysis of the $halo$ region, 
{\it Suzaku} was used because the lowest and stable background are optimum 
to study the faint soft X-ray emission from the halo.
The faint soft X-ray emission from the halo region extends out to $\sim$10 kpc 
perpendicular to the disk, which makes it difficult to estimate the X-ray background emission
from the same {\it Suzaku} FOV.
Therefore, we use an offset observation to evaluate the X-ray background emission.
The offset region covers just next to the $halo$ region of NGC 253.
We also extracted spectra from archival {\it Chandra} data for the three regions.
However, due to lack of sensitivity below 1 keV, the emission lines cannot be clearly detected.
Therefore, we concentrate on {\it XMM-Newton} and {\it Suzaku} data in this paper.
%
%
The sequence numbers, observation dates, pointing directions and exposure times
are summarized in Table \ref{table:observation-log}.

For spectral analysis using {\it XMM-Newton}, 
we removed flaring high background periods in which the count rate is over 0.35 above 10 keV. 
The redistribution matrix files (RMFs) and Ancillary response files (ARFs) were made 
by the Science Analysis System rmfgen and arfgen tools. 
We extracted the background from a source-free region.
In spectra of the $disk$ region, 
two strong instrumental fluorescent emission lines of Al (1.49 keV) and Si (1.74 keV) 
were not subtracted properly.
Since the first line was subtracted too much which gave rise to a large residual around 1.49 keV, 
we ignored the energy band between 1.45 keV and 1.55 keV.
On the other hand, because the second fluorescent line was not subtracted enough, 
we added a gaussian component to make up for this residual fixing the line center at 1.74 keV. 
We did not ignore this energy band in order to evaluate the abundance of Si.
%
%
In spectra of the $superwind$ region, 
no instrumental fluorescent line feature was found 
since the emission from the ISM was much brighter than that of the background.

For spectral analysis of {\it Suzaku}, RMFs were produced by xisrmfgen ftool 
to calculate time-dependent XIS response considering secular degradation 
of the XIS energy resolution.
ARFs were also created 
using xissimarfgen (\cite{arf}). 
For spectral analysis of both the $halo$ region and the offset region, 
we adopted a uniform circle with a radius of 20$'$ as an input image to xissimarfgen.
For the non X-ray background (NXB) component, 
an accumulated dark Earth database with xisnxbgen ftool was used (\cite{xisnxbgen}).
%
%

In this paper, 
a neutral hydrogen column density of the Galaxy is assumed to be 
1.5$\times$10$^{20}$ cm$^{-2}$, based on the LAB survey \citep{lab-survey}.
We assume the solar abundance tabulated in \citet{solar-abundance-table-anders}.
HEAsoft version 6.11 and XSPEC 12.7.0 were utilized.

\begin{table*}[htbp]
\begin{center}
\caption{Observation log}
\label{table:observation-log}
\begin{tabular*}{15cm}{cccccc} \hline\hline
 &                                                    & \multicolumn{2}{c}{{\it Suzaku}}                                                            &  {\it XMM-Newton} \\ \hline         
                                                        &                                                                  & NGC 253                              & NGC 253 OFFSET         &   NGC 253 \\   \hline
Sequence number                      &                                                                   & 805018010                         &  803004010                     &  0152020101\\  
Obs date                                       &                                                                   & 2010/12/14--16                  &  2008/12/29--30              &  2003/06/19--20      \\
FOV center                                   & ($\alpha_{2000}$, $\delta_{2000}$)  &  (00:47:26, -25:13:46)       &  (00:48:22, -25:02:51)    &  (00:47:37, -25:18:10)\\
Exposure                                      & ks                                                              & 101                        & 57                               &  76$^{\ast}$  \\
\hline 
\end{tabular*}
\begin{flushleft} 
\footnotesize{
\hspace{1.cm}$^\ast$ Exposure time after a background flare removal.} \\
\end{flushleft}
\end{center}
\end{table*}

\section{Analysis \& Results}
\label{SEC:analysis-results}
\subsection{Spectral analysis in the $superwind$ region and the $disk$ region}
\label{SEC:innerdisk-disk-analysis}
\begin{figure*}
\begin{minipage}{0.5\hsize}
\begin{center}
\includegraphics[width=8.5cm]{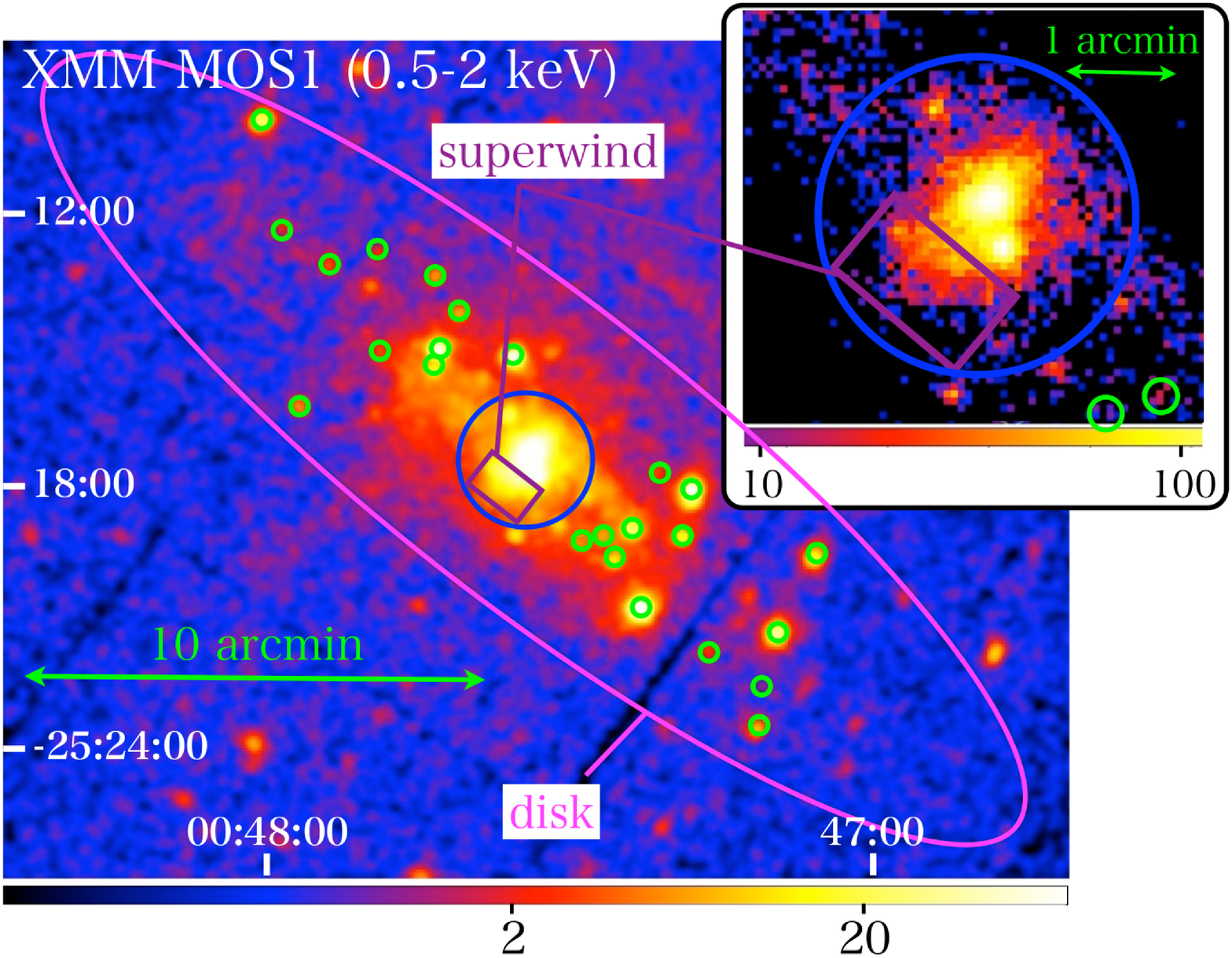} 
   \end{center}
   \end{minipage}  
   \begin{minipage}{0.5\hsize}
\begin{center}
\includegraphics[width=8.5cm]{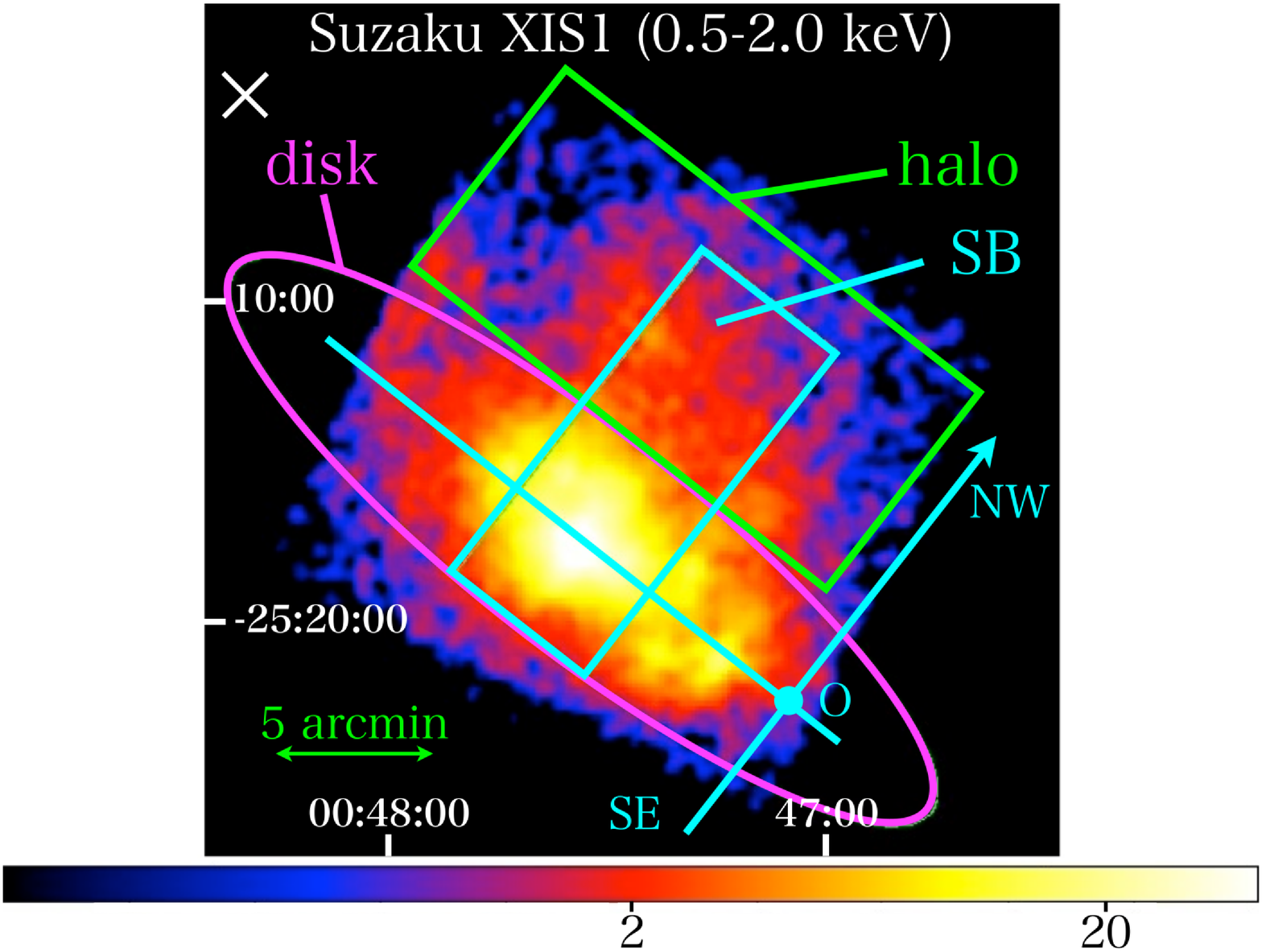}  
\end{center}
 \end{minipage}
 \caption{Left: {\it XMM-Newton} MOS1 image in 0.5--2.0 keV in the unit of cts (76 ks)$^{-1}$ (80 pixel)$^{-1}$. 
 A close-up view of the $superwind$ region is also exhibited in upper right. 
 Purple rectangle and magenta ellipse show the regions to extract abundance patterns 
 denoted as the $superwind$ and $disk$ regions in this paper, respectively.
 Blue and green circles correspond to the excluded regions in spectral analysis of the $disk$ region.
 Right: {\it Suzaku} XIS1 image in 0.5--2 keV in the unit of cts (101 ks)$^{-1}$ (64 pixel)$^{-1}$.
 Green rectangle is the $halo$ region used for the spectral analysis in \S\ref{SEC:spectral-analysis-halo}, and 
 the surface brightness and the hardness ratio extracted from the cyan region denoted as the {\it SB} region 
 are used in \S\ref{SEC:surface-brightness-hardness-ratio}.
 White cross mark shows the aim point of the offset region.
 The images of {\it Suzaku} and {\it XMM-Newton} are smoothed to emphasize extended sources 
 with kernels of $\sigma$ = 25$"$ and 12$"$, respectively but no smoothing is applied for the close-up image.
 The scale is logarithmic. Vignetting and background are not corrected for.
}
 \label{fig:three-regions}
 \end{figure*}

We extracted the spectrum in the $superwind$ and $disk$ regions as shown in Figure \ref{fig:three-regions}.
For spectral analysis of the $disk$ region, 
we removed central 90$"$ region to exclude the $superwind$ region. 
Point sources were also removed with a radius of 10$"$.

Spectra after background subtraction were fitted with emissions 
from absorbed hot interstellar gas and a sum of unresolved point sources.
As absorbers, Galactic and intrinsic absorptions were taken into account.
The former was fixed to the Galactic value while the latter was set to be free.
Optically thin thermal collisionally-ionized equilibrium (CIE) plasma was used to express the hot gas.
Metal abundances in a plasma were divided into five groups, 
i.e., O, Ne, (Mg \& Al), (Si, S, Ar \& Ca) and (Fe \& Ni), based on the metal synthesis mechanism of supernovae.
The abundance of each group is set to be free.
For the hot interstellar gas component, we tried a two-temperature plasma model. 
A one-temperature model cannot reproduce the spectrum due to large residuals 
around emission lines such as O and Fe.
No improvement in the fit was seen when the third-temperature plasma was added.
Metal abundances between the plasmas was linked to each other 
assuming the same origin of the hot gas in the extracted region.
Bremsstrahlung was used to represent the emission from a sum of point sources.
The temperature of the bremsstrahlung was fixed to the typical temperature of 10 keV of the low mass X-ray binary, 
responsible for the emission from the point sources \citep{ngc4631-yamasaki}.
These components were modeled by 
{\it phabs$_{{\rm Galactic}}$} $\times$ {\it phabs$_{{\it superwind~{\rm or}~disk}}$} $\times$ 
({\it vapec$_\mathrm{{\rm 2~{\it T}:~{\it superwind~{\rm or}~disk}}}$} + 
{\it zbremss$_\mathrm{{\it superwind~{\rm or}~disk}}$}) in the XSPEC software.
{\it phabs$_{{\it superwind~{\rm or}~disk}}$}, {\it vapec$_\mathrm{{\rm 2~{\it T}}:~{\it superwind~{\rm or}~disk}}$} and 
{\it zbremss$_\mathrm{{\it superwind~{\rm or}~disk}}$} correspond to 
the absorption column density, a two-temperature thermal plasma and the bremsstrahlung component
in the $superwind$ region or the $disk$ region, respectively.

\begin{figure*}[h!]
  \begin{minipage}{0.5\hsize}
\begin{center}
  \includegraphics[width=8.5cm]{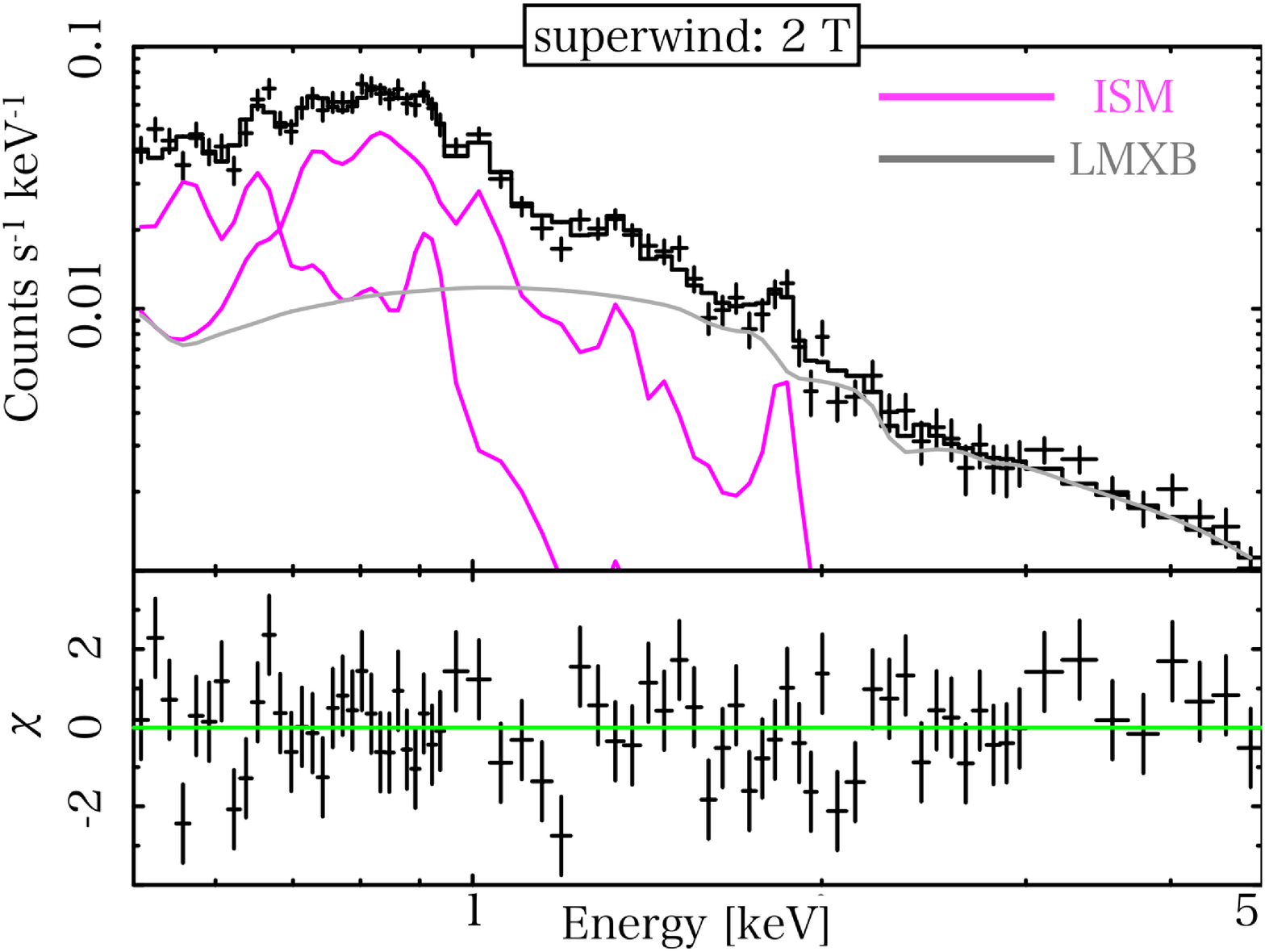} 
   \end{center}
   \end{minipage} 
  \begin{minipage}{0.5\hsize}
\begin{center}
  \includegraphics[width=8.5cm]{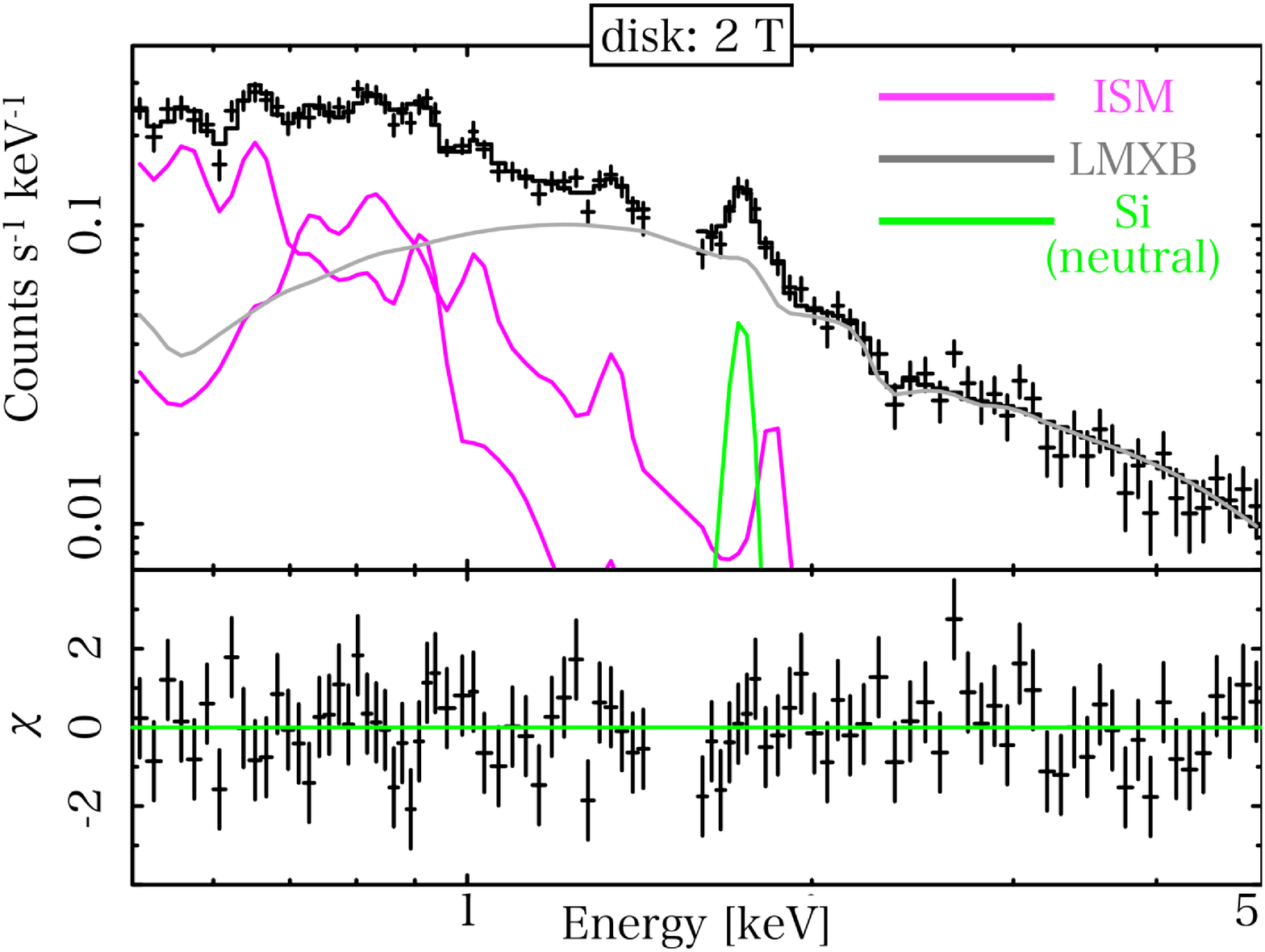} 
   \end{center}
   \end{minipage} 
   \caption{Spectra obtained by {\it XMM-Newton} MOS detector at the energy band of 0.5 and 5 keV 
   in the $superwind$ (left) and $disk$ (right) regions fitted with the two-temperature model.
   Magenta and gray lines correspond to the best fit model for hot interstellar gas and a sum of point sources, respectively.
   We ignored the energy band in 1.45--1.55 keV and added the gaussian component (green) in the spectrum of the $disk$ region 
   due to the detector background of the neutral Al and Si K fluorescent emission lines, respectively.}
\label{fig:xmm-innerdisk-disk-spec-1t-2t}
   \end{figure*}

\begin{figure*}
\begin{center}
\includegraphics[width=12cm]{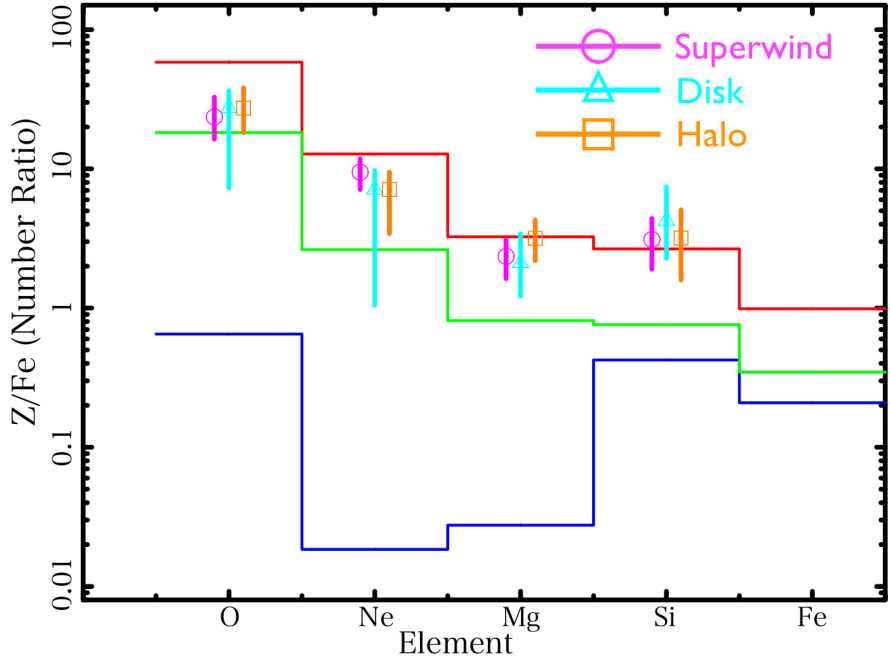}  
\end{center}
 \caption{Abundance patterns of O/Fe, Ne/Fe, Mg/Fe and Si/Fe 
 obtained from spectral analysis in the $superwind$ region (magenta), 
 the $disk$ region (cyan) and the $halo$ region (orange), respectively.
 Blue, green and red lines show expected products by type I supernova \citep{snr-iwamoto}, 
 solar abundance tabulated in \citet{solar-abundance-table-anders} as an example of solar abundance tables 
 and type II supernova \citep{type-ii-nomoto}.}
 \label{fig:abundance-patterns}
 \end{figure*}

Resultant spectra and the best fit parameters are shown 
in Figure \ref{fig:xmm-innerdisk-disk-spec-1t-2t} and summarized in Table \ref{table:innerdisk-disk-best-fit-parameters}.
The best fit temperatures of the two regions are both $\sim$0.2 keV and $\sim$0.6 keV.
For the spectrum of the $disk$ region, 
we ignored the fluorescent Al K$\alpha$ energy band 
and the gaussian component was added to make up for the neutral Si K fluorescent emission line.
We extracted abundance ratios for the best fit model in both regions 
by two-parameter confidence contours between each $\alpha$ element and Fe.
Then, we converted the abundance ratio to a number ratio of the $\alpha$ element to the Fe atom 
using the abundance table that we adopted.
Note that the number ratio is independent of the abundance table 
that we assumed in the fit.
Resulting number ratios are exhibited in Figure \ref{fig:abundance-patterns}.
Abundance patterns of each element in the two regions are consistent.
As comparisons, the number ratios for the solar abundance that we adopted, SN II and SN Ia products 
are also plotted.
For SN II products, 
we referred to those of an average over the Salpeter initial mass function of stellar mass 
from 10 to 50 M$_{\odot}$ in \citet{type-ii-nomoto} with a progenitor metallicity of $Z$ = 0.02 $Z_{\odot}$.
As SN Ia yields, 
we referred to the W7 model in \citet{snr-iwamoto}.
\begin{table}[h!]
\begin{center}
\caption{Best fit parameters for the $superwind$ region and the $disk$ region by {\it XMM-Newton}.}
\label{table:innerdisk-disk-best-fit-parameters}
\begin{tabular}{ccc} \hline\hline
\multicolumn{1}{c}{Region}                                                    & $Superwind$                                       & $Disk$                                              \\\hline
 Model                                                                                        & 2$T$                                                       &  2$T$            \\   \hline
$N_{{\rm H}{~phabs\rm : Galactic}}$ ($\times$10$^{21}$ [cm$^{-2}$])  & 0.15 (fix)                       &  0.15 (fix)         \\
$N_{{\rm H}{~phabs\rm : {\it superwind}~or~{\it disk}}}$ ($\times$10$^{21}$ [cm$^{-2}$])           & $<$0.5             & 1.0$^{+1.4}_{-0.9}$         \\
$kT_{\rm{{\it vapec1} : {\it superwind}~or~{\it disk}}}$ [keV]    & 0.21$\pm$0.02                                    & 0.20$\pm$0.02             \\
O [$Z_{\odot}$]                                                                          & 0.44$^{+0.60}_{-0.20}$                     & 0.28$^{+0.33}_{-0.09}$             \\
Ne [$Z_{\odot}$]                                                                       & 1.3$^{+0.99}_{-0.53}$                        & 0.51$^{+0.79}_{-0.21}$      \\
Mg, Al [$Z_{\odot}$]                                                                 &  1.0$^{+1.1}_{-0.6}$                            & 0.49$^{+0.92}_{-0.26}$               \\
Si, S, Ar, Ca [$Z_{\odot}$]                                                       &  1.4$^{+1.1}_{-0.8}$                           &  1.1$^{+1.6}_{-0.6}$             \\
Fe, Ni [$Z_{\odot}$]                                                                  & 0.35$^{+0.38}_{-0.14}$                     & 0.19$^{+0.22}_{-0.06}$             \\
$\it{Norm_{\rm{{\it vapec1} : {\it superwind}~or~{\it disk}}}}$$^{\ast}$ ($\times$10$^2$)& 7.9$^{+9.3}_{-3.4}$                             & 1.7$^{+4.8}_{-1.4}$                \\
$kT_{\rm{{\it vapec2} : {\it superwind}~or~{\it disk}}}$ [keV]    & 0.62$\pm$0.04                                    & 0.57$^{+0.08}_{-0.21}$                \\
$\it{Norm_{\rm{{\it vapec2} : {\it superwind}~or~{\it disk}}}}$$^{\ast}$ ($\times$10$^2$) &  6.6$^{+3.8}_{-3.4}$                            & 0.5$^{+0.8}_{-0.3}$                \\
O/Fe$^{\dagger}$                                                                    & 24$^{+9}_{-7}$                                     & 27$^{+9}_{-20}$                \\
Ne/Fe$^{\dagger}$                                                                  &  9.5$\pm2.4$                                        &  7.1$^{+2.6}_{-6.1}$               \\
Mg/Fe$^{\dagger}$                                                                  & 2.4$\pm0.7$                                         & 2.1$^{+1.3}_{-0.9}$               \\
Si/Fe$^{\dagger}$                                                                    & 3.1$^{+1.3}_{-1.2}$                             & 4.3$^{+3.2}_{-2.0}$                 \\ \hline
$kT_{\rm{{\it zbremss}:{\it superwind~{\rm or}~disk}}}$$^{\ddagger}$ [keV] & 10 (fix)                          & 10 (fix)                \\
Flux$_{\rm{{\it zbremss} : {\it superwind}~or~{\it disk}}}$$^{\S}$~[$\times$10$^{-14}$~erg s$^{-1}$ cm$^{-2}$]  &8.4$\pm{0.5}$ & 90$^{+6}_{-4}$  \\
$\chi^2/d.o.f$                                                                            & 90/63                                                      & 83/79               \\ 
\hline 
\end{tabular}
\end{center}
\begin{flushleft} 
\scriptsize{
\hspace{2.5cm}$^\ast$ Normalization of the $\it{apec}$ model divided by a solid angle $\Omega$, 
assumed in a uniform-sky ARF calculation (20' radius), \\
\hspace{2.7cm}i.e. $\it{Norm} = (1/\Omega) \int n_e n_H dV / (4\pi(1+z)^2)D^2_A)$ cm$^{-5}$ sr$^{-1}$ 
in unit of 10$^{-14}$, where $D_A$ is the angular diameter distance. \\
\hspace{2.5cm}$^\dagger$ Number ratio relative to the Fe atom obtained from two-dimensional contour maps between Z and Fe. \\
\hspace{2.5cm}$^{\ddagger}$ Temperature of the zbremss component.\\
\hspace{2.5cm}$^\S$ Unabsorbed flux in the extracted region in 0.5--2 keV.} 
\end{flushleft}
\end{table}
\subsection{Spectral analysis in the halo region}
\label{SEC:spectral-analysis-halo}

\subsubsection{Evaluation of the X-ray background emission}
For spectral analysis of the $halo$ region, 
{\it Suzaku} data were utilized.
Firstly, we analyzed the data of the offset region to evaluate the X-ray background emission.
The offset region is located just next to NGC 253 as shown in Figure \ref{fig:three-regions}.
We removed point-source candidates and 
diffuse-like sources by a visual inspection and a wavelet algorithm (by a CIAO tool wavdetect) in 0.4--2.0 keV.
\begin{figure*}
\begin{center}
\includegraphics[height=6cm]{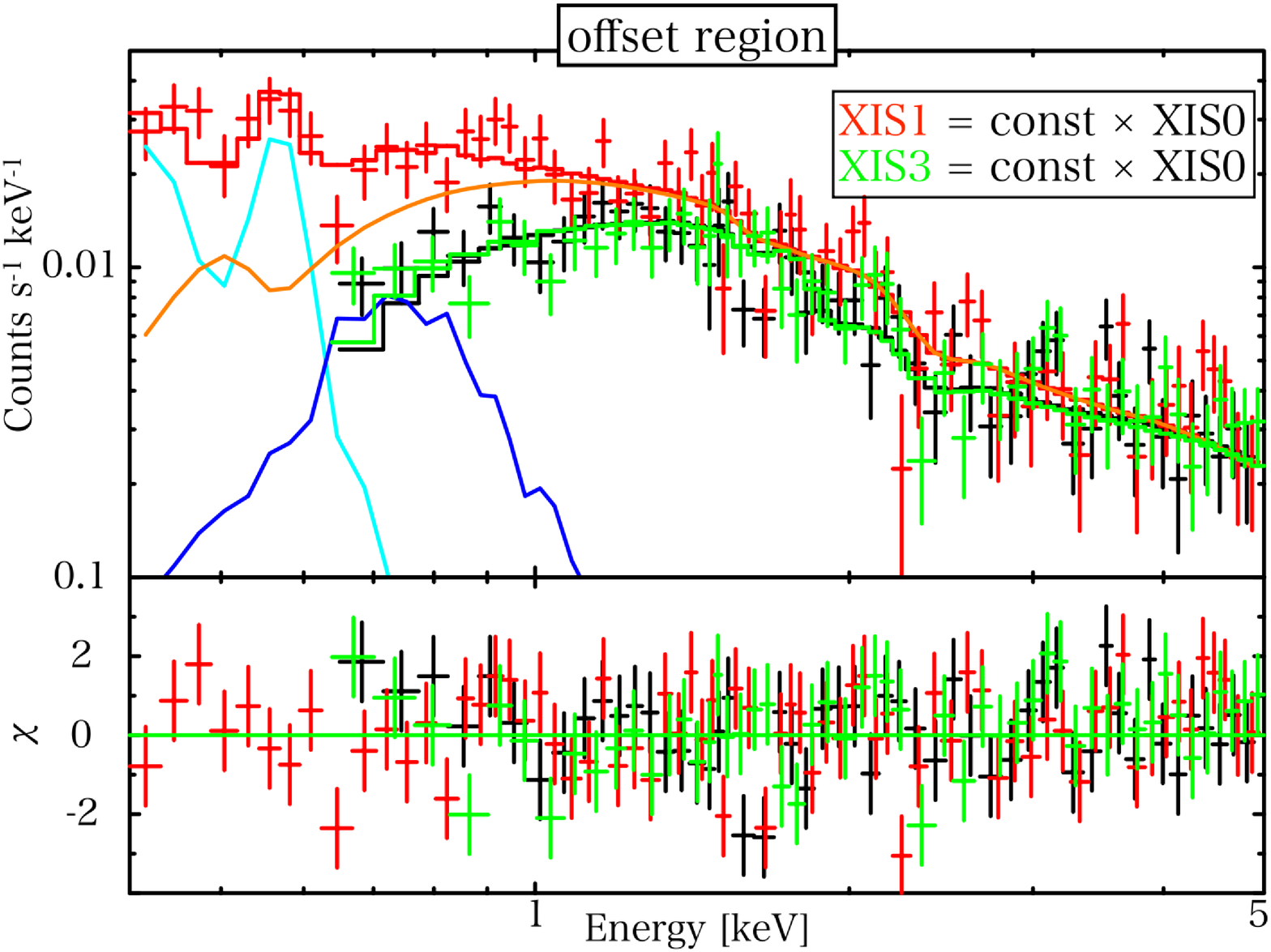}
\end{center}
   \caption{Spectrum obtained in the offset region fitted with the typical X-ray background emission model, 
   i.e., $\it{apec_{\rm{SWCX+LHB}}}$ (cyan) + $\it{phabs_{\rm{Galactic}}}\times(\it{apec_{\rm{MWH}}}~({\rm blue})$ + {\it power-law}$_{\rm CXB}$ (orange)).
   Three components correspond to SWCX+LHB, MWH and CXB, respectively. 
   For simplicity, the best fit model is shown for only the XIS1 detector.
   For spectra of the XIS1 and XIS3 detectors, the constant factor is multiplied to consider the systematic error 
   in different sensors.}
\label{fig:offset-04-2keV-xis1-image}
   \end{figure*}
We fitted the spectrum with the typical X-ray background emission expressed by a sum of 
(1) an unabsorbed thin thermal CIE plasma, 
(2) an absorbed thin thermal CIE plasma and (3) an absorbed power law.
The first two components represent emissions from the local area 
consisting of solar wind charge exchange (SWCX: e.g., \citet{2004ApJ...610.1182S}) 
and Local Hot Bubble (LHB) and 
Galactic halo (Milky Way Halo: MWH).
On the other hand, the last component corresponds to the accumulation of unresolved extragalactic point sources 
(cosmic X-ray background: CXB) described by an absorbed power-law model with a photon index of 1.4 
as shown in \citet{cxb-kushino}.
Thus, we adopted the following models in the XSPEC software:
{\it apec$_\mathrm{SWCX+LHB}$} + {\it phabs$_{{\rm Galactic}}$}$\times$ 
({\it apec$_\mathrm{MWH}$}+ {\it power-law}$_{{\rm CXB}}$).
We carried out a simultaneous fitting using the XIS0, 1 and 3 detectors at the energy range of 
0.4--5.0 keV for the BI (XIS1) and 0.6--5.0 keV for the FIs (XIS0 and XIS3).
For the XIS1 and XIS3 detectors, a constant factor denoted as {\it f} was multiplied 
to consider the systematic error in different sensors.
Resulting spectra and parameters are shown in Figure \ref{fig:offset-04-2keV-xis1-image} 
and Table \ref{table:suzaku-xis0-1-3-spec_05-5keV_offset-halo_1t-2t}.
Resultant temperatures of the Galactic foreground and 
the surface brightness of the CXB component are consistent with typical values 
obtained in previous studies (e.g., \citet{cxb-kushino,yoshino-san}).
The data from the offset observations is also analyzed in \citet{2010PASJ...62.1423S}, 
which shows the best-fit parameters consistent with our results within the statistical errors.

\subsubsection{Estimation of the stray light from the inner regions}
\label{SEC:morekomi-estimation}
Due to the poor {\it Suzaku} angular resolution (HPD$\sim$1.8$'$), 
a stray light from the inner region has to be evaluated 
to derive physical properties of the hot ISM in the $halo$ region properly.
As the origin of the stray light, emissions from not only the $disk$ region but also the nuclear region were taken into account 
because the nuclear region is extremely bright in hard X-ray due to the most intense starburst activity
\citep{2011ApJ...742L..31M}.
As a stray light source, we firstly considered the emission from the nuclear region with a radius of 20$''$  
based on \citet{2011ApJ...742L..31M}.
The estimated count rate from the nuclear region is $\sim$3\% compared to NXB subtracted spectra 
in the $halo$ region at the energy range between 0.5 and 5 keV.
Therefore we concluded that the stray light from the nuclear region is negligible.
Next, we extracted the whole disk region excluding the nuclear region by {\it XMM-Newton}.
The spectrum was modeled well by two thin thermal plasmas and a bremsstrahlung with $\chi^2/d.o.f$ = 271/250.
The bremsstrahlung component from point sources dominates even in the soft band and 
metal abundances can not be constrained.
Therefore, metal abundances of two thin thermal plasmas were set to be 1 solar.
The estimated count rate from the stray light is $\sim$20 \% compared to NXB subtracted spectra 
in the $halo$ region in 0.5--5 keV.
%

Thus, as the stray light source, these models were taken into account 
fixing all parameters to the best fit values in spectral analysis of the $halo$ region.

\subsubsection{Spectral fitting for the halo region}
\label{SEC:spectral-fitting}
Spectra of the $halo$ region were fitted with the X-ray background, hot interstellar gas (1 $T$ or 2 $T$),  
a sum of point sources and the stray light from the disk modeled by 
{\it apec$_\mathrm{SWCX+LHB}$} + {\it phabs$_{{\rm Galactic}}$} $\times$ ({\it apec$_\mathrm{MWH}$} + {\it power-law}$_{{\rm CXB}}$) + 
{\it phabs$_{{\rm Galactic}}$} $\times$ {\it phabs$_{{\it halo}}$} $\times$ ({\it vapec$_\mathrm{{\rm 1~{\it T}~or~2~{\it T}:~{\it halo}}}$} + 
{\it zbremss$_\mathrm{{\it halo}}$}) 
+ {\it phabs$_{{\rm Galactic}}$} $\times$ {\it phabs$_{{\it disk}}$} $\times$ ({\it apec$_\mathrm{{\rm 2~{\it T}}:~{\it disk}}$} + 
{\it zbremss$_\mathrm{{\it disk}}$}).
Parameters on the X-ray background were linked between NGC 253 and the offset region and 
consistent with those of the obtained only in the offset region within the statistical error.
The fourth terms indicate the stray-light from the inner disk region.

\begin{figure*}
\begin{minipage}{0.5\hsize}
\begin{center}
\includegraphics[width=8.5cm]{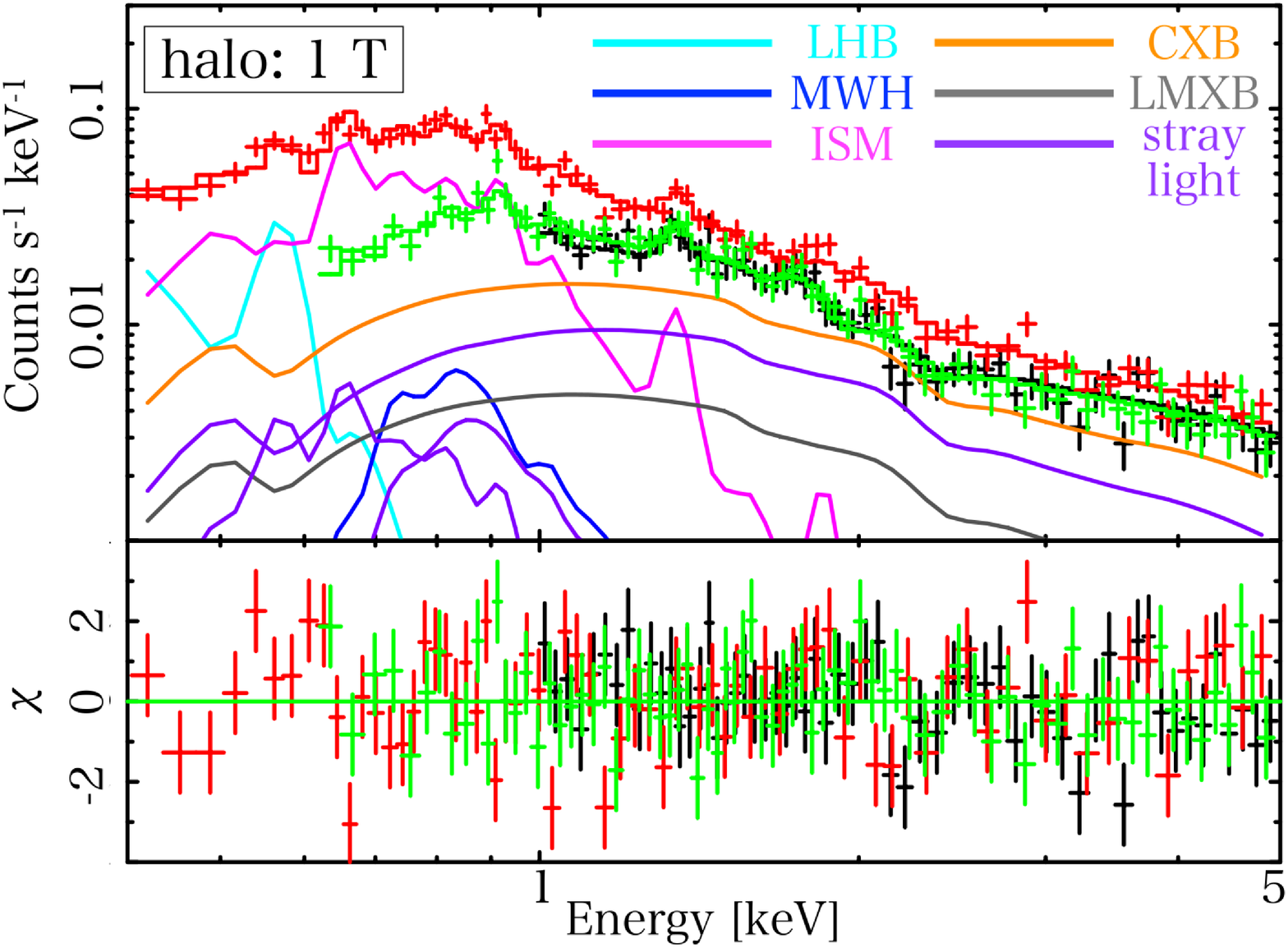} 
   \end{center}
   \end{minipage}  
   \begin{minipage}{0.5\hsize}
\begin{center}
\includegraphics[width=8.5cm]{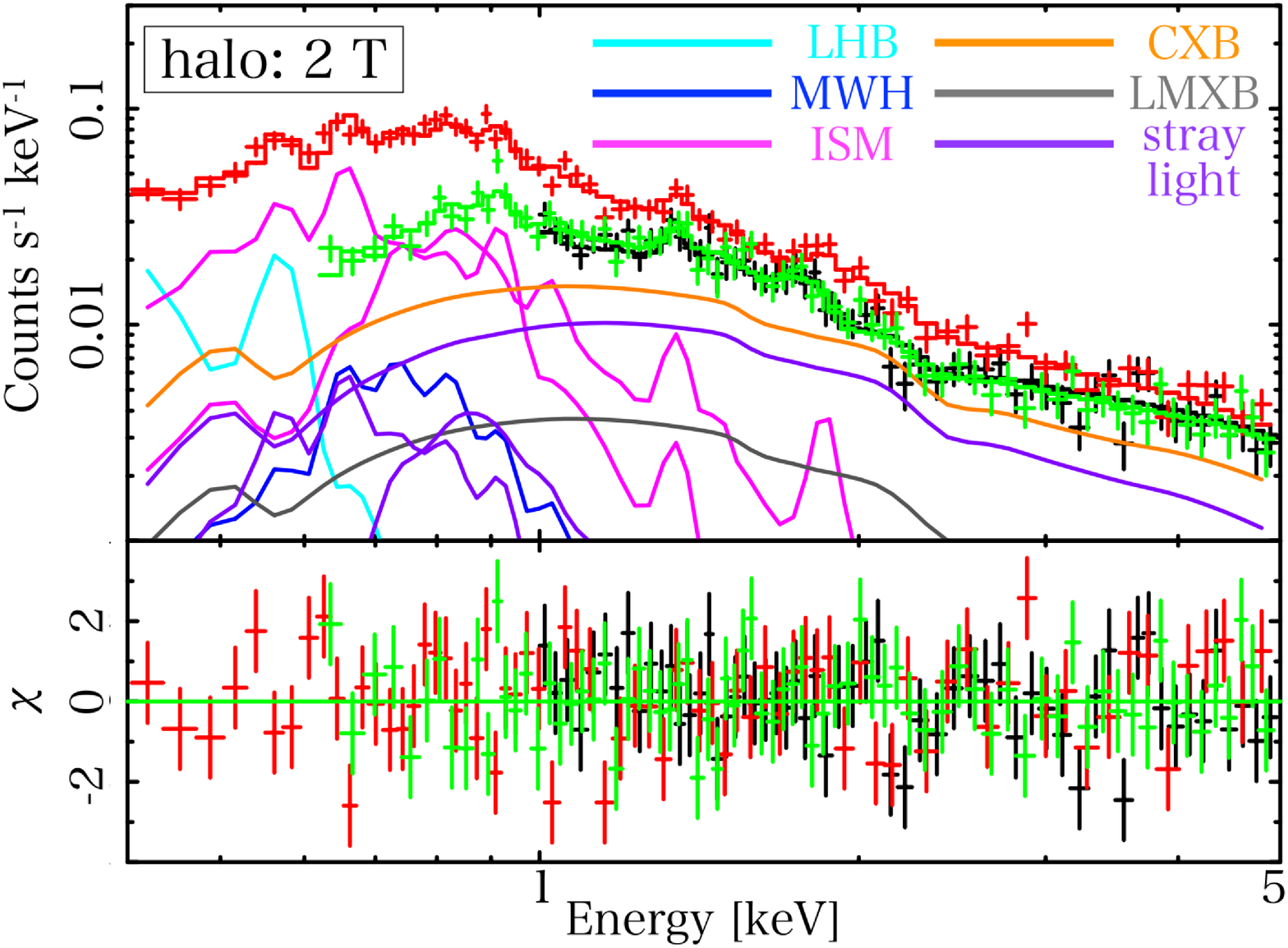}  
\end{center}
 \end{minipage}
 \caption{NXB-subtracted XIS0 (black), XIS1 (red) and XIS3 (green) spectra observed by {\it Suzaku} 
 at the energy range between 0.4 and 5.0 keV in the $halo$ region. 
 The lines show the best-fit model that consists of 
 {\it apec$_\mathrm{SWCX+LHB}$} (cyan), {\it phabs$_{{\rm Galactic}}$} $\times$ {\it apec$_\mathrm{MWH}$} (blue), 
   {\it phabs$_{{\rm Galactic}}$} $\times$ {\it power-law}$_{{\rm CXB}}$ (orange), 
   {\it phabs$_{{\rm Galactic}}$} $\times$ {\it phabs$_\mathrm{{\it halo}}$} $\times$ 
   ({\it vapec$_\mathrm{{\rm 1~{\it T}~or~2~{\it T}:~{\it halo}}}$ + {\it zbremss$_\mathrm{{\it halo}}$}}) 
   (magenta and gray) and {\it phabs$_{{\rm Galactic}}$} $\times$ {\it phabs$_\mathrm{{\it disk}}$} $\times$ 
   ({\it apec$_\mathrm{2~{\it T}:~{\it disk}}$} + {\it zbremss$_\mathrm{{\it disk}}$}) (purple). 
   The first three terms show the X-ray background which is composed of SWCX+LHB, MWH and CXB components and 
   the forth and fifth terms indicate the ISM and the sum of point sources in the $halo$ region.
   The last term corresponds to the stray light from the inner disk region.
   The X-ray background was linked between the offset region and the $halo$ region.
   As the ISM component, one- (left) and two-temperature (right) models are adopted, respectively. 
}
 \label{fig:suzaku-xis0-1-3-spec_05-5keV_halo-1t-2t}
 \end{figure*}

Resultant spectra and the best fit parameters are shown in Figure \ref{fig:suzaku-xis0-1-3-spec_05-5keV_halo-1t-2t} 
and Table \ref{table:suzaku-xis0-1-3-spec_05-5keV_offset-halo_1t-2t}.
The goodness of the fit improved slightly by adding the second thermal component.
The third thermal component was not required. 
The resultant temperature of the one-temperature model is almost consistent with 
that of the obtained by {\it XMM-Newton} \citep{ngc253-xmm-halo} 
while temperatures of the two-temperature model are about twice higher than those of \citet{ngc253-xmm-halo}.
This discrepancy may be caused by the difference of the adopted energy band for spectral analysis.
In \citet{ngc253-xmm-halo}, they adopted the energy band of 0.2--1.5 keV and 
no line feature of Mg and Si was observed owing to the lack of photons.
As indicated in Figure \ref{fig:suzaku-xis0-1-3-spec_05-5keV_halo-1t-2t}, 
the higher temperature plasma with k$T$ $\sim$0.6 keV is responsible for emission lines 
such as Ne\emissiontype{X}, Mg\emissiontype{XI} and Si\emissiontype{XIII}.
Abundance patterns for one- and two-temperature models were extracted 
as is the case of the $superwind$ region and the $disk$ region as shown in Table \ref{table:suzaku-xis0-1-3-spec_05-5keV_offset-halo_1t-2t}.
Abundance patterns for both models are consistent with each other within the statistical error.
Resulting number ratios indicate that rich $\alpha$ elements exist even in the $halo$ region, 
which suggests that metals in the $halo$ region are provided 
by Type II supernovae associated with the inner starburst activity.
\renewcommand{\baselinestretch}{1.0}\selectfont
\begin{table}[h!]
\begin{center}
\caption{Best fit parameters for the offset region and the $halo$ region by {\it Suzaku}.}
\label{table:suzaku-xis0-1-3-spec_05-5keV_offset-halo_1t-2t}
\begin{tabular}{cccc} \hline\hline
\multicolumn{1}{c}{Region}                                        &   OFFSET                                             & \multicolumn{2}{c}{$Halo$}                                 \\\hline
 Model                                                                            &                                                               & 1$T$                                                                & 2$T$                                                       \\   \hline
{\it f}~$^{\ast}$ (XIS1)                                                  & 1.02$^{+0.08}_{-0.07}$                     & 1.12$^{+0.07}_{-0.06}$                               & 1.08$^{+0.05}_{-0.04}$                                      \\
{\it f}~$^{\ast}$ (XIS3)                                                  & 0.95$^{+0.08}_{-0.07}$                     & 0.93$^{+0.06}_{-0.05}$                              & 0.94$\pm$0.04                                      \\
$N_{{\rm H}{~phabs\rm : Galactic}}$ ($\times$10$^{21}$ [cm$^{-2}$])  & 0.15 (fix)                                 &  $\leftarrow$                                                 & $\leftarrow$         \\
$kT_{apec : \rm{SWCX+LHB}}$ [keV]                                    & 0.07$^{+0.02}_{-0.01}$                     &  0.09$^{+0.01}_{-0.02}$                              & 0.07$^{+0.02}_{-0.01}$                       \\
$\it{Norm_{apec : \rm{SWCX+LHB}}}$$^{\dagger}$~/~{\it f} & 227$^{+996}_{-166}$                        &  84$^{+169}_{-31}$                                     & 242$^{+912}_{-172}$                          \\    
$kT_{apec : \rm{MWH}}$ [keV]                                  & 0.29$^{+0.10}_{-0.06}$                     & 0.62$^{+0.13}_{-0.41}$                               & 0.29$^{+0.10}_{-0.06}$                       \\
$Norm_{apec : \rm{MWH}}$$^{\dagger}$~/~{\it f} & 1.7$^{+1.1}_{-0.7}$                            & 0.75$\pm$0.26                                             & 1.8$^{+1.1}_{-0.7}$                               \\
$\Gamma$$_{power-law : {\rm CXB}}$ (fix)           &  1.4 (fix)                                                 & $\leftarrow$                                                   & $\leftarrow$                                            \\
$\it{SB}$$_{power-law : {\rm CXB}}$$^\ddagger$& 8.7$\pm$0.5                                        & 8.6$\pm$0.3                                                  & 8.7$\pm$0.3                               \\ \hline
$N_{{\rm H}{~phabs\rm : {\it disk}}}$ ($\times$10$^{21}$ [cm$^{-2}$])&                             & 0.5 (fix)                                                           & $\leftarrow$                                              \\ 
$kT_{apec : \rm{{\it disk}1}}$ [keV]                            &                                                               &  0.21 (fix)                                                        & $\leftarrow$                                              \\
$\it{Norm_{apec : \rm{{\it disk}1}}}$$^{\dagger}$~/~{\it f} &                                                    &   2.2 (fix)                                                          &  $\leftarrow$                                            \\    
$kT_{apec : \rm{{\it disk}2}}$ [keV]                            &                                                               &   0.73 (fix)                                                        &  $\leftarrow$                                             \\
$\it{Norm_{apec : \rm{{\it disk}2}}}$$^{\dagger}$~/~{\it f}&                                                     &  0.53 (fix)                                                         &  $\leftarrow$                                             \\  
$kT_{\rm{{\it zbremss}:{\it disk}}}$$^{\S}$ [keV]       &                                                              & 10 (fix)                                                              & $\leftarrow$                                               \\
Flux$_{\rm{{\it zbremss}:{\it disk}}}$$^{\sharp}$~/~{\it f}~[$\times$10$^{-13}$~erg s$^{-1}$ cm$^{-2}$]   &&1.3 (fix)                              & $\leftarrow$                                              \\  \hline
$N_{{\rm H}{~phabs\rm : {\it halo}}}$ ($\times$10$^{21}$ [cm$^{-2}$]) &                           & $<$0.5                                                             & $<$0.7                                                      \\
$kT_{\rm{vapec1}} : {\it halo}$ [keV]                            &                                                           & 0.30$\pm$0.02                                              & 0.22$\pm$0.02                         \\
O [$Z_{\odot}$]                                                            &                                                                 & 0.45$^{+1.12}_{-0.21}$                                & 0.43$^{+0.28}_{-0.14}$                         \\
Ne [$Z_{\odot}$]                                                          &                                                                 & 0.60$^{+1.43}_{-0.27}$                                & 0.84$^{+0.74}_{-0.37}$                        \\
Mg, Al [$Z_{\odot}$]                                                    &                                                                 & 1.1$^{+0.94}_{-0.62}$                                    &  1.2$^{+1.2}_{-0.6}$                              \\
Si, S, Ar, Ca [$Z_{\odot}$]                                          &                                                                & 1.2$^{+5.4}_{-1.2}$                                       &  1.3$^{+1.8}_{-0.7}$                               \\
Fe, Ni [$Z_{\odot}$]                                                     &                                                                & 0.22$^{+0.52}_{-0.13}$                                & 0.29$^{+0.23}_{-0.11}$                         \\
$\it{Norm_{\rm{vapec1} : {\it halo}}}$$^{\dagger}$~/~{\it f}   &                                               & 43$^{+38}_{-29}$                                          &  36$^{+46}_{-13}$                                   \\
$kT_{\rm{vapec2} : {\it halo}}$ [keV]                       &                                                                 &                                                                           & 0.59$^{+0.19}_{-0.11}$                          \\
$\it{Norm_{\rm{vapec2} : {\it halo}}}$$^{\dagger}$~/~{\it f}   &                                               &                                                                           &  11$^{+9.1}_{-5.1}$                                \\
O/Fe$^{\flat}$                                                              &                                                                 & 35$^{+13}_{-9}$                                            & 27$^{+11}_{-9}$                         \\
Ne/Fe$^{\flat}$                                                            &                                                                 & 6.8$^{+1.8}_{-1.3}$                                      & 7.1$^{+2.4}_{-3.7}$                        \\
Mg/Fe$^{\flat}$                                                            &                                                                 & 3.8$\pm$1.2                                                   &  3.2$^{+1.1}_{-1.0}$                              \\
Si/Fe$^{\flat}$                                                              &                                                                & 3.5$^{+3.4}_{-3.5}$                                       &  3.2$^{+1.9}_{-1.6}$                               \\
$kT_{\rm{{\it zbremss}:{\it halo}}}$$^{\S}$ [keV]    &                                                                 & 10 (fix)                                                              & $\leftarrow$                                               \\
Flux$_{\rm{{\it zbremss}:{\it halo}}}$$^{\sharp}$~/~{\it f}~[$\times$10$^{-14}$~erg s$^{-1}$ cm$^{-2}$]  &     &6.4$^{+1.9}_{-1.8}$     & 4.8$^{+2.1}_{-2.0}$ \\
$\chi^2/d.o.f$                                                              & 194/169                                                 & 434/367                                                             & 413/365                                                     \\ 
\hline 
\end{tabular}
\end{center}
\begin{flushleft} 
\scriptsize{
\hspace{2.3cm}$^\ast$ Constant factor relative to the XIS0 detector.\\
\hspace{2.3cm}$^\dagger$ Normalization of the $\it{apec}$ model divided by a solid angle $\Omega$, 
assumed in a uniform-sky ARF calculation (20' radius), \\
\hspace{2.5cm}i.e. $\it{Norm} = (1/\Omega) \int n_e n_H dV / (4\pi(1+z)^2)D^2_A)$ cm$^{-5}$ sr$^{-1}$ 
in unit of 10$^{-14}$, where $D_A$ is the angular diameter distance. \\
\hspace{2.3cm}$^\ddagger$ Surface brightness of the $power-law$ model in the unit of 
photons s$^{-1}$ cm$^{-2}$ sr$^{-1}$ keV$^{-1}$ at 1 keV\@. \\
\hspace{2.3cm}$^\S$ Temperature of the zbremss component. \\
\hspace{2.3cm}$^{\sharp}$ Unabsorbed flux in the extracted region in 0.5--2 keV. \\
\hspace{2.3cm}$^{\flat}$ Number ratio relative to the Fe atom obtained from two-dimensional contour maps between Z and Fe.} 
\end{flushleft}
\end{table}
\renewcommand{\baselinestretch}{1.5}\selectfont
\subsection{Surface brightness and hardness ratio}
\label{SEC:surface-brightness-hardness-ratio}
In this section, we investigated physical properties of the hot ISM 
such as density and temperature.
To infer these physical parameters, we extracted surface brightness and hardness ratio.  
The soft band below 1 keV was used where the ISM component dominates 
as shown in Figure \ref{fig:suzaku-xis0-1-3-spec_05-5keV_halo-1t-2t}.
We extracted the central bright rectangle region of 5.2$'$$\times$13$'$ 
(hereafter the {\it SB} region) along with the minor axis perpendicular to the disk ($<$ 3 kpc from the center) 
toward the halo (3--10 kpc from the center)
as shown in Figure \ref{fig:three-regions} colored with cyan 
where higher S/N ratios were expected.

\begin{figure*}[h!]
\centerline{
\includegraphics[width=8cm]{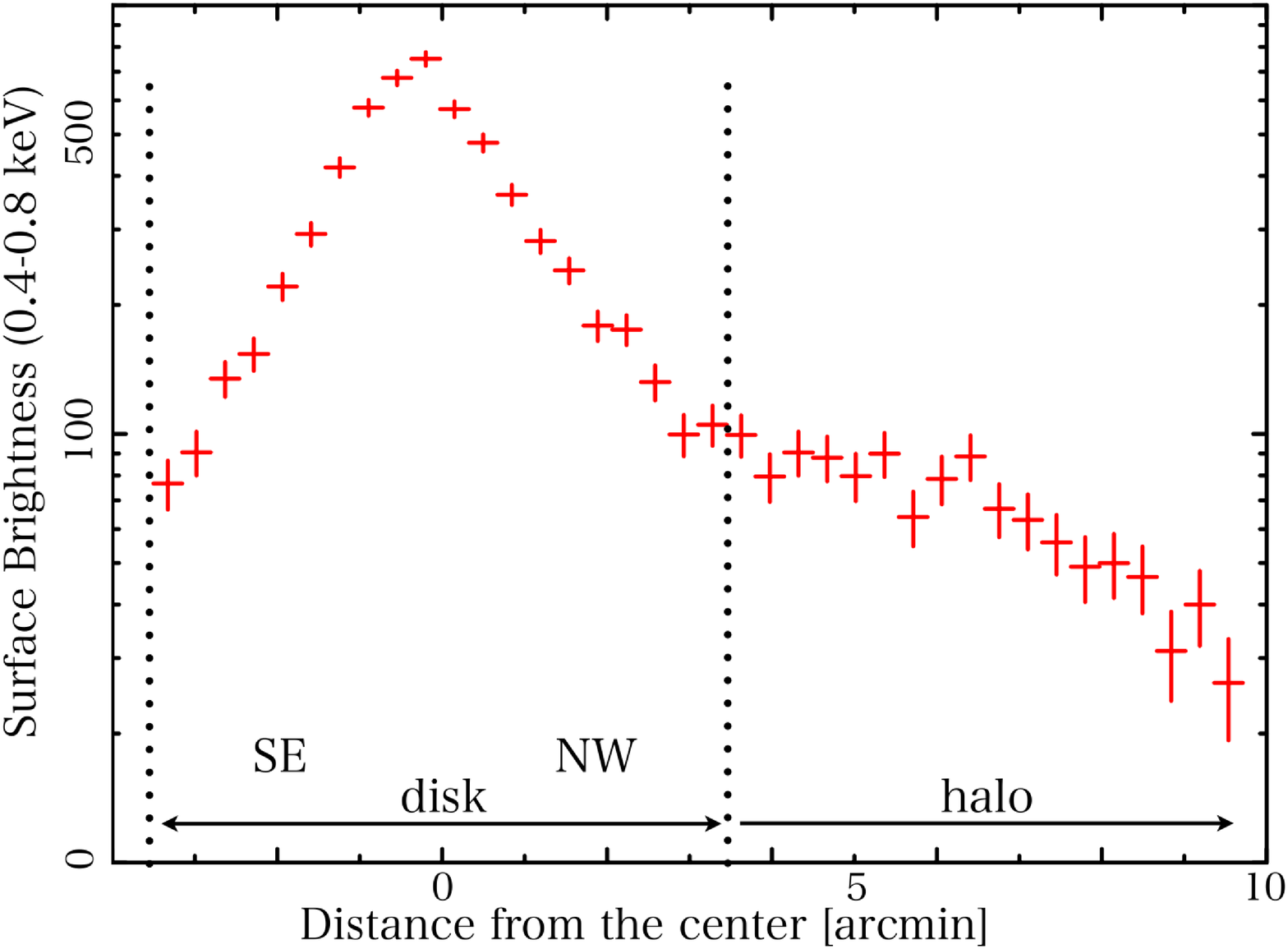}
	\hspace{0.3cm}
\includegraphics[width=8cm]{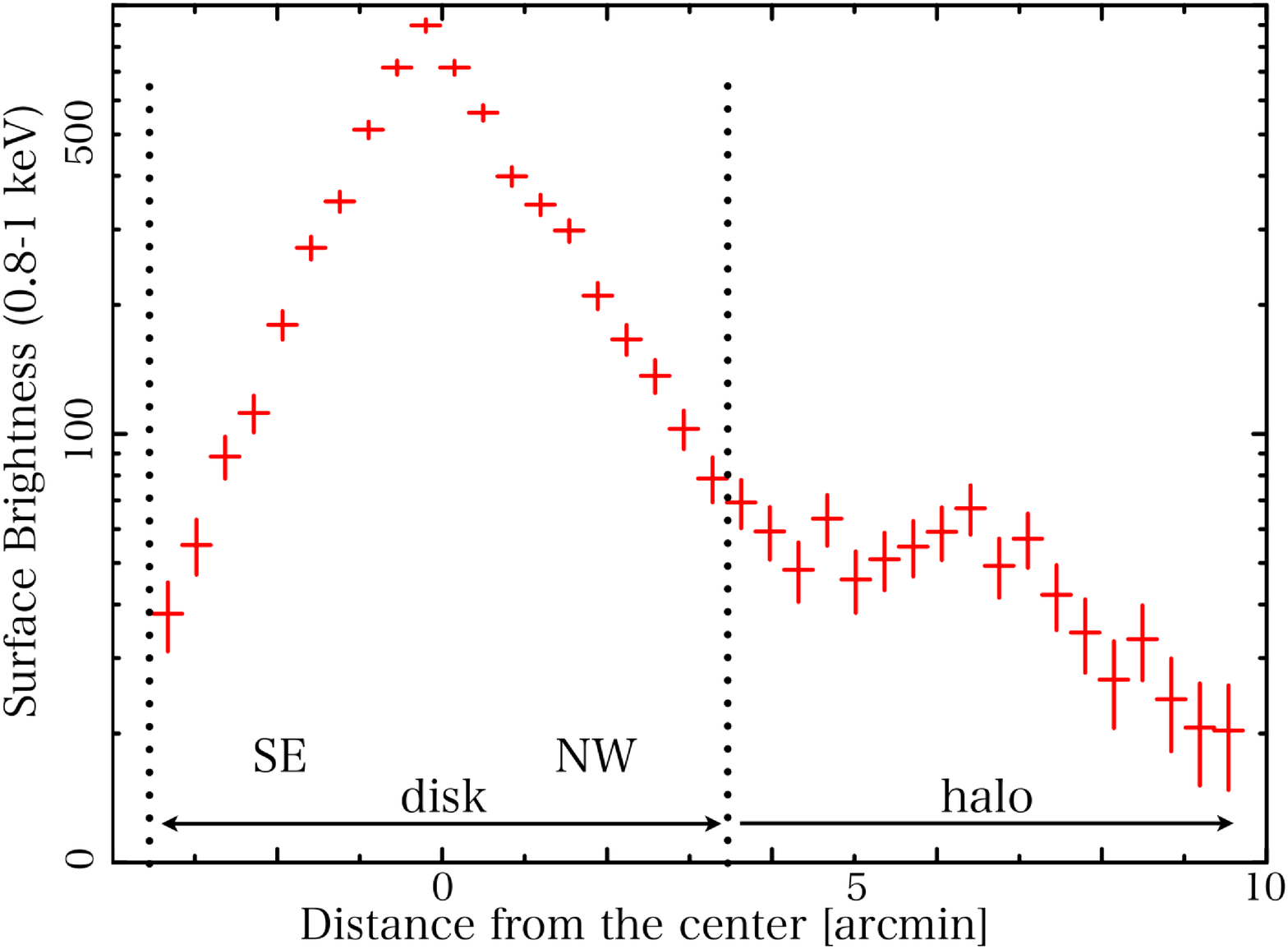}
}
\vspace{0.3cm}
\centerline{
\includegraphics[width=8cm]{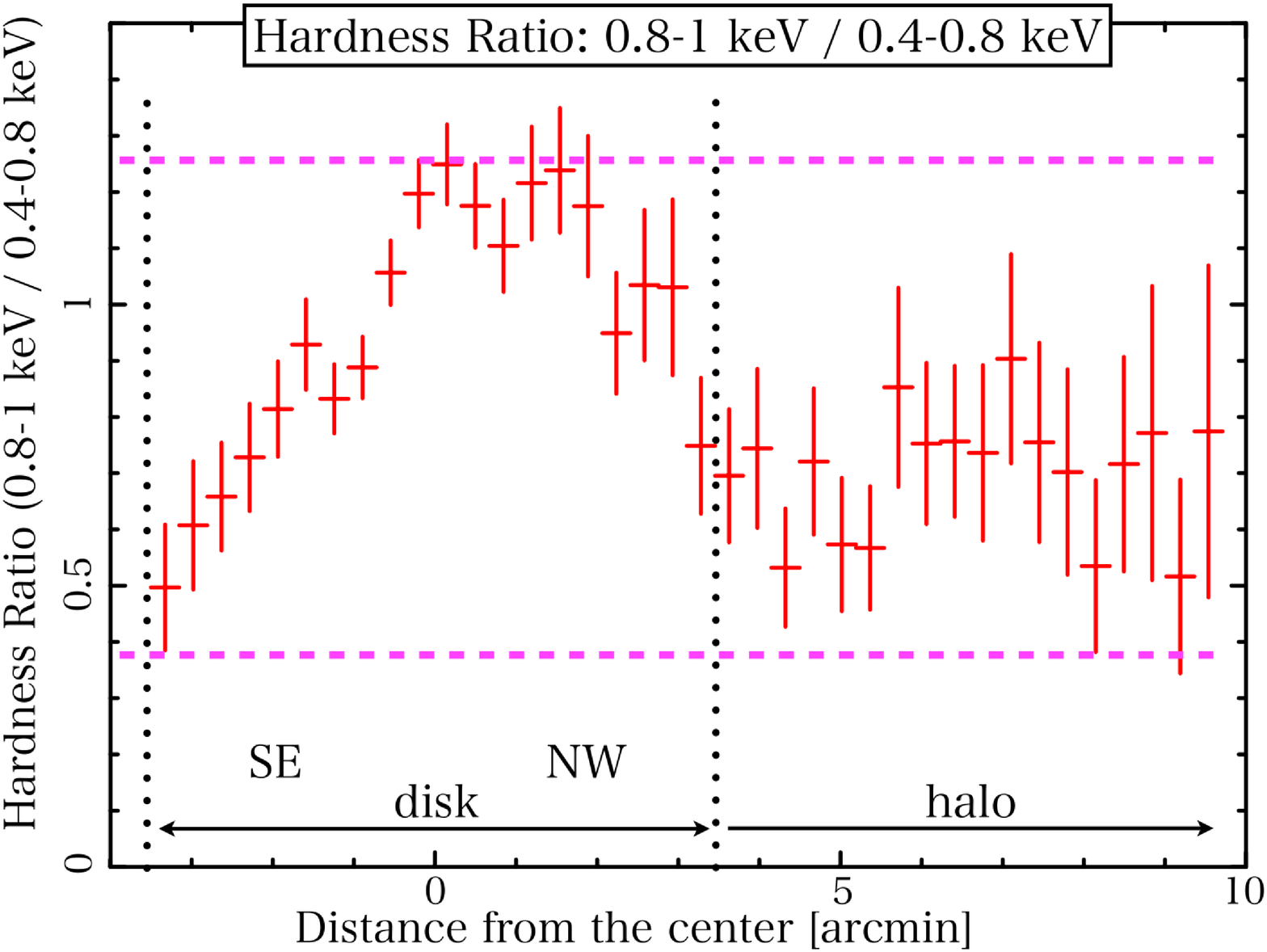}
}
   \caption{Surface brightness after a background removal in the {\it SB} region at the energy range 
   between 0.4 and 0.8 keV (top left) and 0.8 and 1.0 keV (top right) 
   along with the minor axis perpendicular to the disk.
   Hardness ratio of 0.8--1.0 keV to 0.4--0.8 keV (bottom)  
   was derived by dividing surface brightness profiles of each energy band.
   Dashed and dotted lines indicate corresponding temperatures of 0.2 and 0.6 keV 
   assuming one-temperature thin thermal plasma 
   with the absorption column density of N$_{H}$ = 5$\times$10$^{20}$ cm$^{-2}$ and 
   the border between the disk and the halo, respectively.}
\label{fig:sb-hr-reg2-3-04-08keV-08-1keV}
   \end{figure*}

We extracted the surface brightness in two energy bands: 0.4--0.8 keV as the soft band and 
0.8--1.0 keV as the hard band.
The two energy bands are selected so that roughly the same number of photons are included.
A contribution of the X-ray background, estimated from the simulation, was subtracted.
To evaluate the contribution of the Galactic and CXB components, 
an observation of the background sky was simulated with the xissim ftool.
We input the best fit parameters of spectral analysis obtained in the offset region 
(see second column of Table \ref{table:suzaku-xis0-1-3-spec_05-5keV_offset-halo_1t-2t}) into the simulation.
The statistical error associated with the simulation is $<$3\% and hence negligible.
For the estimation of the NXB component, we utilized the xisnxbgen ftool.
These procedures allow us to extract information of the density and temperature 
of the ISM itself accurately.
Then, for the Galactic and CXB components, a vignetting correction was carried out 
assuming a monochromatic energy in each energy band; 0.6 keV for the soft band 
and 0.9 keV for the hard band.
We confirmed that this energy selection did not change our results significantly.
Resultant surface brightness profiles after a background removal are shown in Figure \ref{fig:sb-hr-reg2-3-04-08keV-08-1keV} top. 
Its behavior seems to be different between the disk and the halo in the {\it SB} region.
A slope in the disk is steeper than that of the halo in 0.4--0.8 keV 
while a bump-like structure is seen in the halo in 0.8--1.0 keV.
To extract a hardness ratio profile, 
the surface brightness profile in 0.8--1 keV was divided by that in 0.4--0.8 keV.
Resulting hardness ratio profile is shown in Figure \ref{fig:sb-hr-reg2-3-04-08keV-08-1keV} bottom.
A hardness ratio gradient is found in the disk 
while no hardness ratio gradient is observed in the halo.
This suggests that a temperature gradient may exist only in the disk.
To ensure this suggestion, an absorption effect 
that changes apparent hardness has to be taken into account.
%
The neutral hydrogen gas is widely distributed from the central region to the halo \citep{ngc253-hi}.
Its column density decreases gradually towards outside the disk from $\sim$5$\times$10$^{21}$ cm$^{-2}$ around the central region 
to $\sim$5$\times$10$^{20}$ cm$^{-2}$ around the border between the disk and the halo.
Thus, the column density in the halo is estimated to be $<$5$\times$10$^{20}$ cm$^{-2}$.
\citet{ngc253-xmm-halo} shows that significant extra absorption was not required in the SE disk and 
the halo of their paper, 
while additional absorption on the order of 10$^{21}$ cm$^{-2}$ was required in several regions of the NW disk.
We also obtained using {\it XMM-Newton} that the 90 \% upper limit of the absorption column density is 5$\times$10$^{20}$ cm$^{-2}$
in the region corresponding to 2--5 bin in extracted profiles of the NW disk in the {\it SB} region.
Variation on the hardness value due to extra absorption is within 5 \% up to the column density of 1$\times$10$^{21}$ cm$^{-2}$. 
Therefore, the gradient and the flatness on the hardness ratio profile is likely due to temperature difference.
Only for the first bin of the NW disk in the {\it SB} region from the center may absorption affect the hardness ratio 
due to the densest molecular cloud on the order of 10$^{23\mathchar`-24}$ cm$^{-2}$ \citep{ngc253-radio-sakamoto}.
\section{Discussion}
\subsection{Comparison with other starburst galaxies on abundance patterns}
\label{SEC:comparison-other-starburst-galaxies-abundance-patterns}
\begin{figure*}
\begin{center}
  \includegraphics[height=8cm]{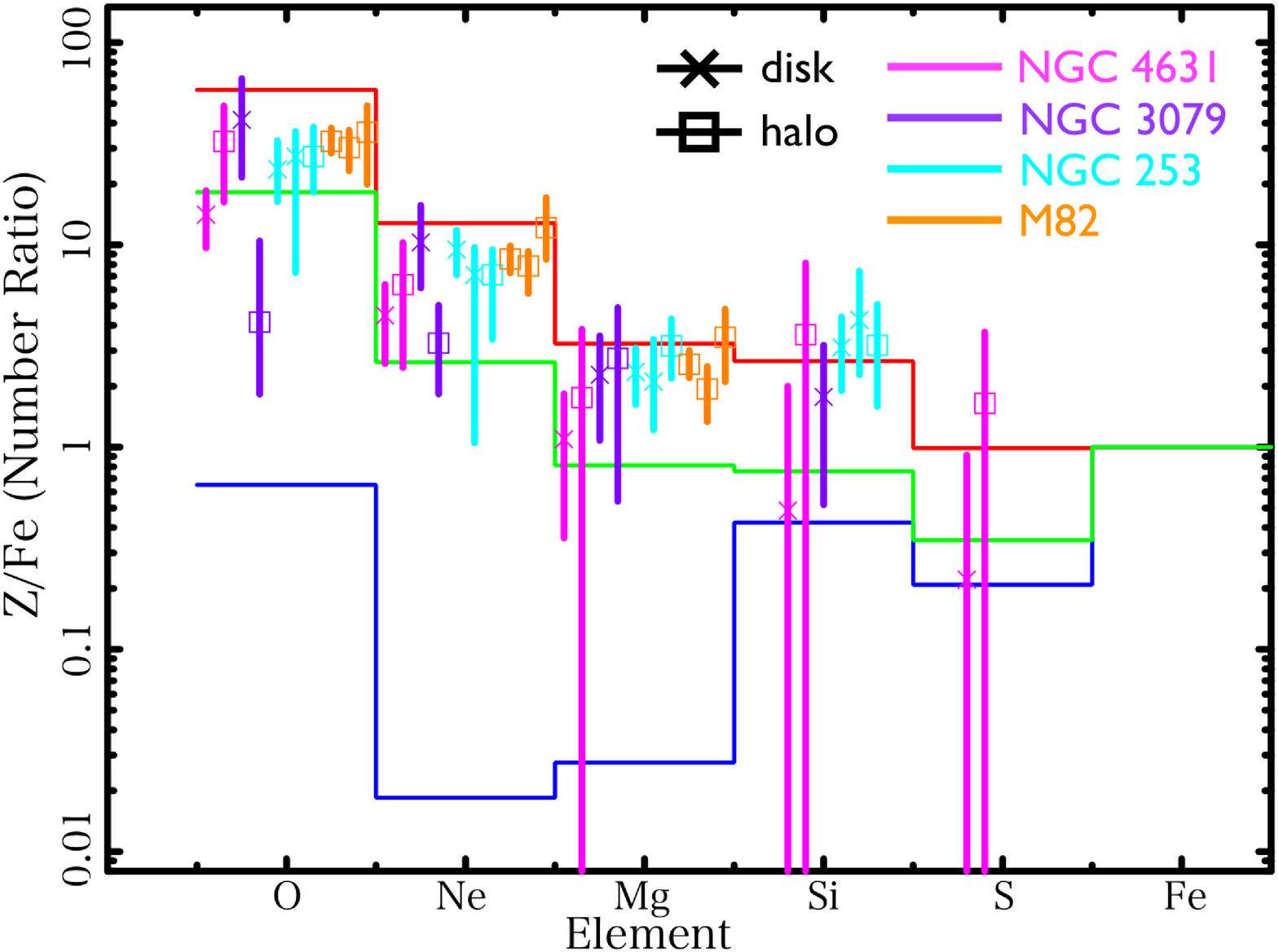} 
   \end{center}
   \caption{Abundance patterns for starburst galaxies of 
   NGC 4631 (magenta), NGC 3079 (purple), NGC253 (cyan) and M82 (orange), respectively.
   Cross and rectangle marks indicate the disk region and the halo region, respectively.
   For NGC 3079, we referred to abundance patterns of the central 0.5$'$ circle and the 1$'$-2$'$ ring region 
   in \citet{2012arXiv1205.6005K} as the disk and halo regions, respectively.}
\label{fig:suzaku-halo-abundance-patters-m82-ngc4631-ngc253}
   \end{figure*}

In \S\ref{SEC:innerdisk-disk-analysis} and \S\ref{SEC:spectral-analysis-halo}, 
abundance patterns for three regions in NGC 253 are obtained.
In this section, we compare abundance patterns of NGC 253 
with those of other starburst galaxies, i.e., NGC 4631, M82 and NGC 3079,  
taken from the statistically best values of \citet{ngc4631-yamasaki}, \citet{2011PASJ...63S.913K} and \citet{2012arXiv1205.6005K}.
A comparison of abundance patterns is shown in Figure \ref{fig:suzaku-halo-abundance-patters-m82-ngc4631-ngc253}.
Abundance patterns in the disk region of four starburst galaxies are consistent with one another and 
are heavily contaminated by SN II, probably associated with the starburst activity in the central region.
Furthermore, abundance patterns of NGC 253 in the $halo$ region are also SN II -like and 
consistent with those of NGC4631 and M82 in the halo, which suggests that 
the same chemical pollution mechanism works and that hot gas in the halo is provided 
by the central starburst region.
In the case of NGC 3079, the abundance pattern in the halo is closer to that of solar. 
This result means that pollution may not be sufficient yet  
because NGC 3079 is considered to be in an early phase of the starburst \citep{2012arXiv1205.6005K}.
\subsection{Energetics of the halo region}
\label{SEC:energetics}
We discuss the possibility 
that hot gas in the $halo$ region may be energetically provided from the inner starburst region.
In the $halo$ region, observed unabsorbed luminosity and thermal energy ($3nkTV\eta/2$) of the hot gas 
in 0.5--10 keV are (3.8$^{+2.4}_{-0.5}$ and 2.3$^{+1.0}_{-0.7}$)$\times$10$^{38}$ erg s$^{-1}$ and 
(2.2$^{+0.8}_{-0.6}$ and 3.3$^{+2.5}_{-1.3}$)$\times$10$^{55}$ $\eta^{1/2}$ erg 
for low and high temperature plasmas 
in the best fit two-temperature model as shown in 
Table \ref{table:suzaku-xis0-1-3-spec_05-5keV_offset-halo_1t-2t}.
We adopted a sum of the luminosity and the thermal energy of each plasma 
as the total observed luminosity and thermal energy.
$\eta$ ($<$1) corresponds to the volume filling factor of the X-ray emitting gas 
as shown in \citet{2002ApJ...576..745C} and 
the same volume filling factor is assumed for both plasmas.
The density is proportional to $\eta^{-1/2}$ and 
therefore the thermal energy is proportional to $\eta^{1/2}$.
The lower limit of supernova rate around the nuclear region of NGC 253 
are on the order of 0.1 yr$^{-1}$ \citep{2006AJ....132.1333L}.
Assuming the typical starburst duration of 10$^{6\mathchar`-7}$ yr, 
luminosity and thermal energy of a supernova 
of 10$^{36\mathchar`-37}$ erg s$^{-1}$ and 10$^{51\mathchar`-52}$ erg \citep{lmc-snrs}, 
the expected total luminosity and thermal energy from supernovae during a starburst duration are 
calculated to be 10$^{41\mathchar`-43}$ erg s$^{-1}$ and 10$^{56\mathchar`-58}$ erg, respectively.
Considering the total observed luminosity and thermal energy in the $halo$ region, 
the energy of the observed thermal plasma may be supplied from the nuclear region 
if 0.01--50 $\eta^{1/2}$ \% of the total emission in the nuclear region has flowed to the $halo$ region.
Note that from this discussion alone, 
we cannot constrain the contribution of other possible origins discussed in 
e.g., \citet{2009AN....330.1028H}, \citet{2011ApJ...736L..27S} and \citet{2011A&A...535A..79H}.
\subsection{Dynamics of hot interstellar gas}
\label{SEC:dynamics-of-the-hos-ism-gas}
\begin{figure*}[h!]
\centerline{
\includegraphics[width=8cm]{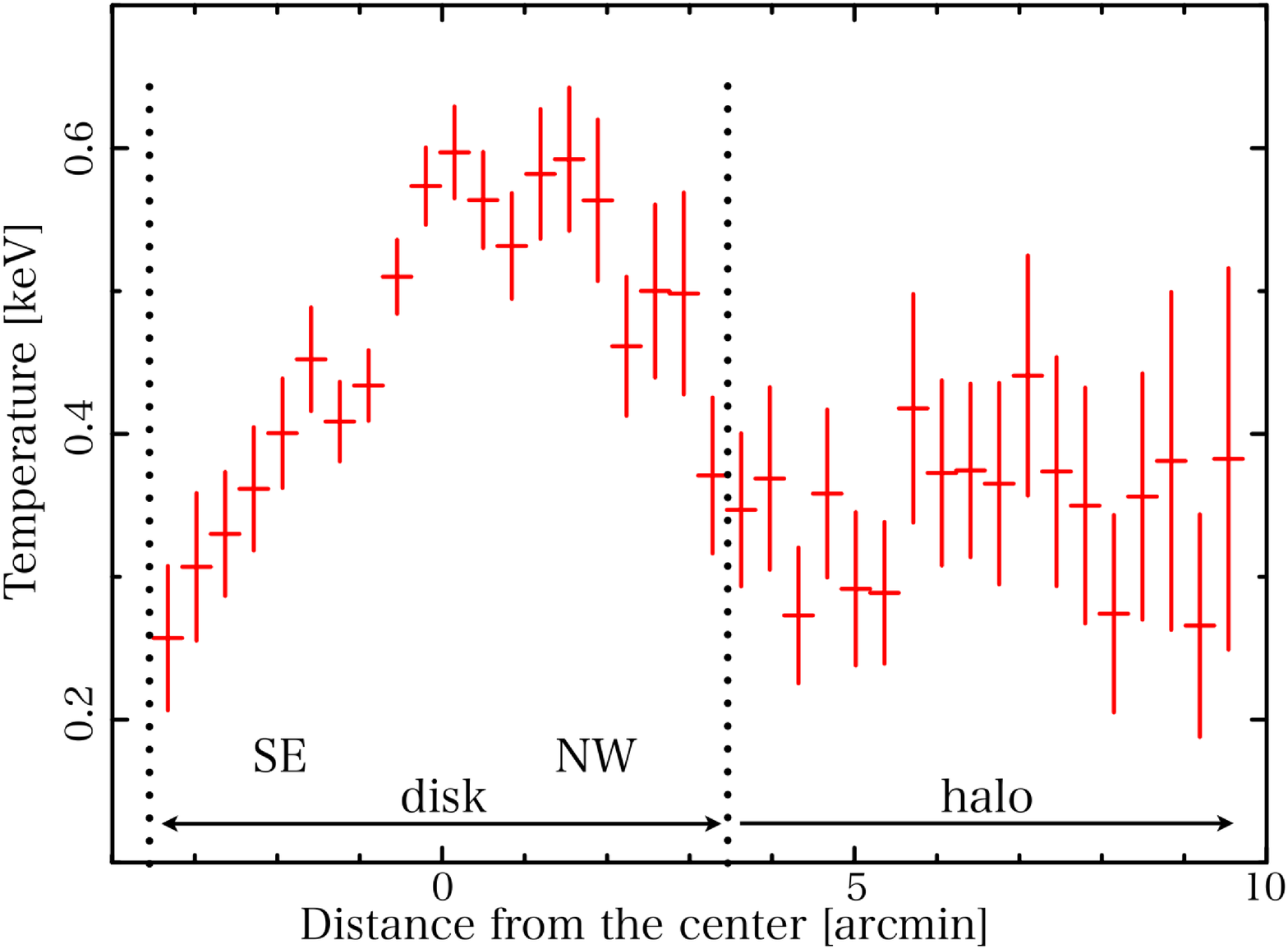}
}
\vspace{0.3cm}
\centerline{
\includegraphics[width=8cm]{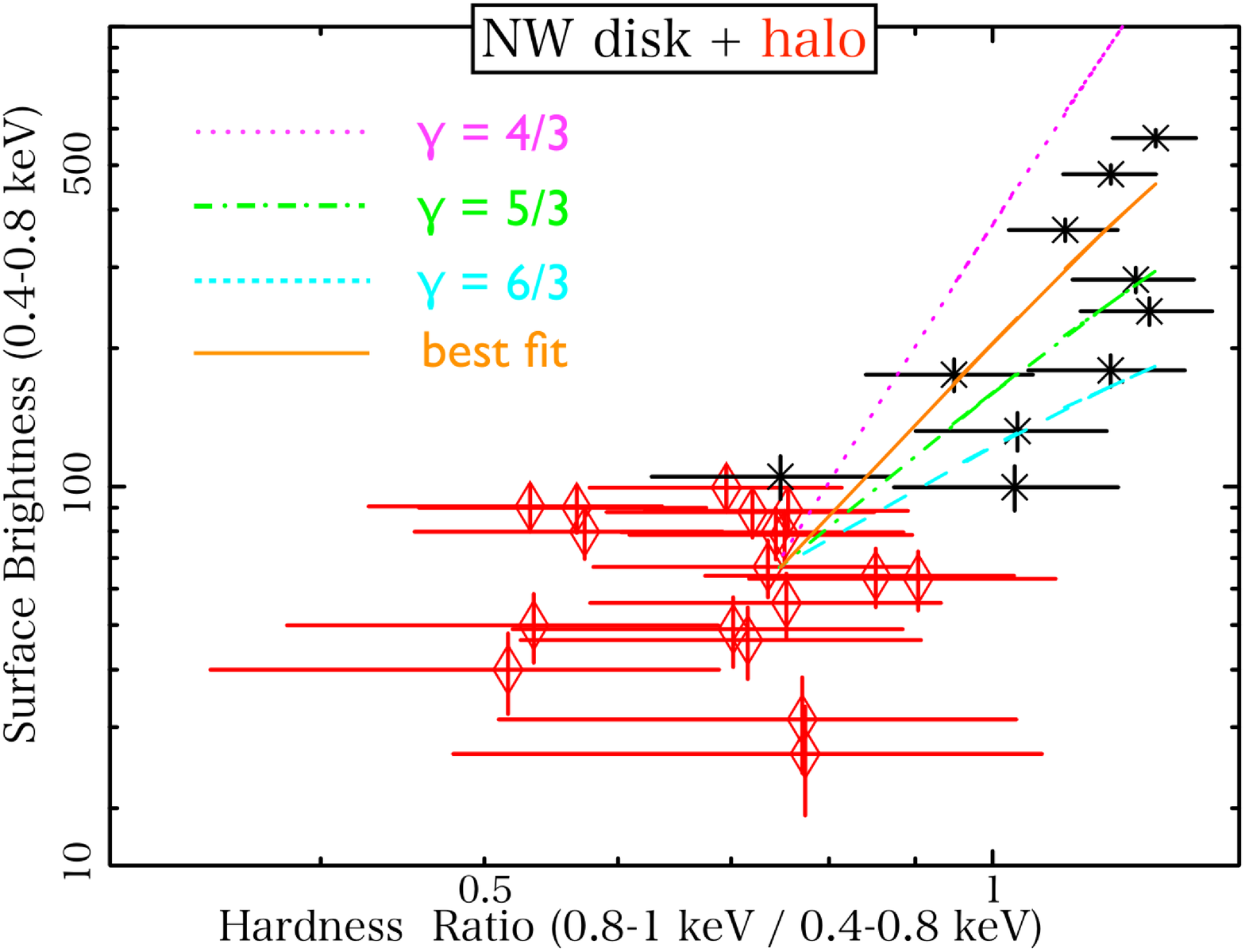}
	\hspace{0.3cm}
\includegraphics[width=8cm]{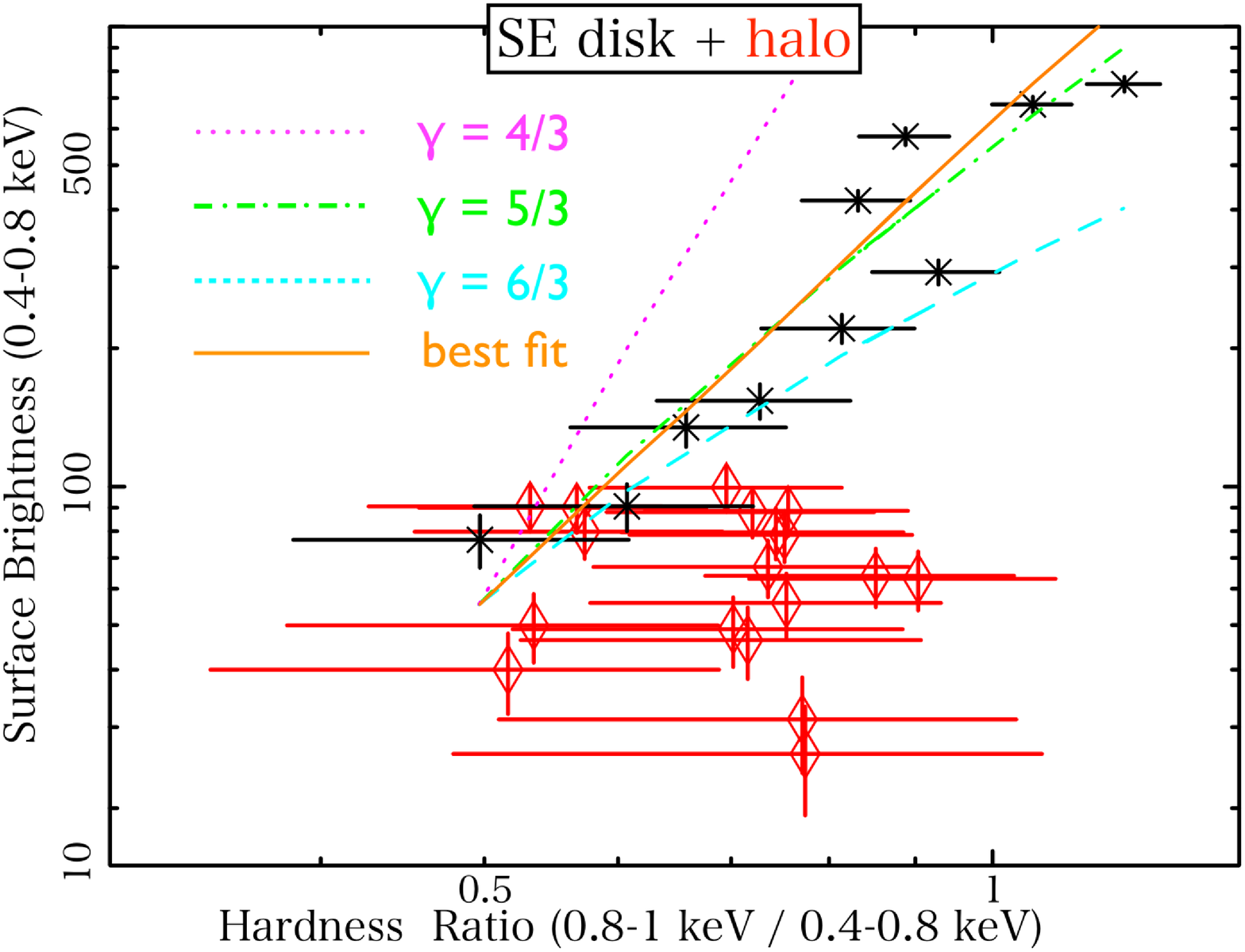}
}
 \caption{Top: temperature profile obtained from the hardness ratio profile in the {\it SB} region 
 assuming the simplest model of the absorbed thin thermal plasma with the column density of 5$\times$10$^{20}$ cm$^{-2}$.
 Hardness ratio vs the surface brightness at the energy range 
 between 0.4 and 0.8 keV in the NW disk (bottom left) and the SE disk (bottom right). 
 Cross and diamond marks show results in the disk and the halo, respectively. 
 As reference values, polytropic indices of 6/3 (dash), 5/3 (dot-dash) and 4/3 (dot) 
 and the best fir curves (solid) between density and temperature are also plotted together.
 The normalization of each curve is arbitrarily determined.
}
 \label{fig:temp-profile-hr-vs-sb}
 \end{figure*}

In \S\ref{SEC:surface-brightness-hardness-ratio}, 
surface brightness and hardness ratio profiles in the {\it SB} region were obtained.
In order to examine dynamics of hot interstellar gas, 
we extracted information on the density and temperature in the disk and the halo of NGC 253 
from the profiles.
Firstly, we derived the temperature profile from the hardness ratio profile as shown in Figure \ref{fig:temp-profile-hr-vs-sb} top.
The hardness ratio is converted to the temperature assuming a thin thermal plasma model 
({\it vapec} in the XSPEC software) multiplied by the absorption column density of 5$\times$10$^{20}$ cm$^{-2}$ 
convolved with the $Suzaku$ response matrix.
Metal abundances are fixed to the best fit values in the best fit model 
as shown in Table \ref{table:suzaku-xis0-1-3-spec_05-5keV_offset-halo_1t-2t}.
The temperature in the halo of the {\it SB} region is about 0.3 keV which is consistent with the obtained 
by the one-temperature model of spectral analysis in the $halo$ region as indicated 
in Table \ref{table:suzaku-xis0-1-3-spec_05-5keV_offset-halo_1t-2t}.
The temperature ranges from 0.2 keV to 0.6 keV.
These temperatures are also consistent with the obtained by the two-temperature model in spectral analysis
which reflect typical temperatures with differential emission measure in the $disk$ region and the $halo$ region.
Thus, it is indicated that the resultant temperature profile traces the typical temperature of the ISM 
under the assumption of the one-temperature model.

Next, information on the density was extracted.
The surface brightness can be expressed by 
${\it \Lambda}$($T$) $n_{{\rm e}}$$n_{{\rm H}}$$V$ $\sim$ ${\it \Lambda}$($T$) $n_{{\rm H}}$$^2$$V$, 
where ${\it \Lambda}$($T$), $n_{{\rm e}}$, $n_{{\rm H}}$ and $V$ indicate emissivity, electron and hydrogen densities and volume, respectively.
To calculate the volume we need the length of the hot gas in line-of-sight direction.
Since we do not know the actual size, here we simply assume that the line-of-sight length is the same 
in the extracted {\it SB} region.
In this case, the surface brightness of each projected region is proportional to the density squared 
and the emissivity (dependent on temperature).
Here we assume that the volume filling factor is unity.
We extracted the emissivity in 0.4--0.8 keV as is the case with \citet{2007ApJ...669..158G} 
using XSPEC assuming the temperature and abundances obtained 
from the hardness ratio and spectral fitting 
as shown in Table \ref{table:innerdisk-disk-best-fit-parameters} and \ref{table:suzaku-xis0-1-3-spec_05-5keV_offset-halo_1t-2t}, respectively.
Finally, we plotted the hardness ratio and the surface brightness as shown in Figure \ref{fig:temp-profile-hr-vs-sb} bottom. 
As reference values, polytropic indices ($T$$\rho^{1-\gamma}$ = const) of 6/3, 5/3 and 4/3 
between the density and temperature are plotted together.
The observed gradient between the hardness ratio and the surface brightness can be expressed 
by the curve with the polytropic index of 5/3 for both the NW disk and the SE disk.
We confirmed that the best-fit polytropic indices are 
1.5$^{+0.1}_{-0.4}$ and 1.6$^{+0.2}_{-0.3}$ for the NW disk region and the SE disk region, respectively.
Equation of state of hot gas with polytropic index of 5/3 corresponds to be adiabatic.
In contrast, the surface brightness declines gradually in the halo of the {\it SB} region 
with a shallower slope compared to that of the disk keeping hardness ratios constant.
This suggests that the hot interstellar gas expands from the inner disk adiabatically towards the halo and 
then expands freely from the inner part of the halo towards the outer part of the halo as the outflow.
\subsection{Constraint on the velocity of the outflow in the halo}
\label{SEC:velocity-constraint}
\begin{figure*}
\begin{center}
  \includegraphics[height=8cm]{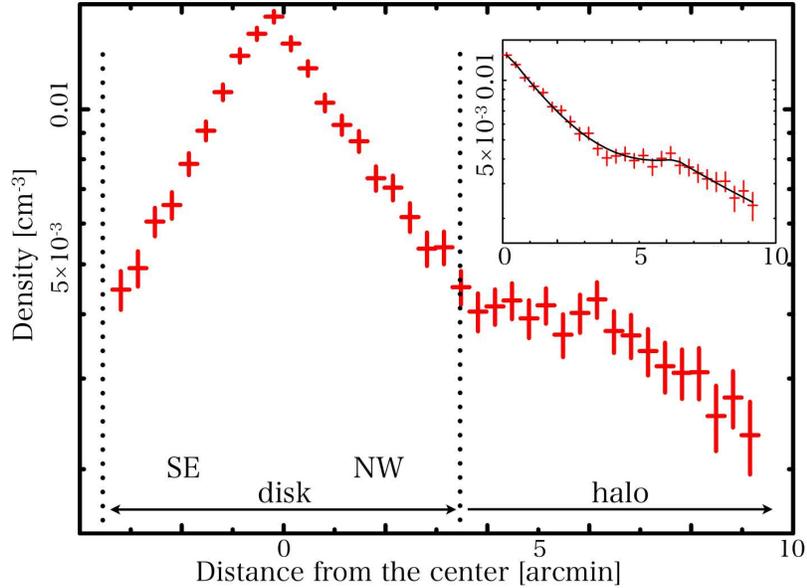} 
   \end{center}
   \caption{Density profile estimated by the surface brightness profile in the {\it SB} region 
   at the energy range between 0.4 and 0.8 keV 
   as shown in Figure \ref{fig:sb-hr-reg2-3-04-08keV-08-1keV} top left.
   Upper right in this figure shows the density profile in the NW disk and the halo 
   fitted with the double exponential function.}
\label{fig:suzaku-xis1-04-08keV-density-profile}
   \end{figure*}

\begin{figure*}
\begin{center}
  \includegraphics[height=8cm]{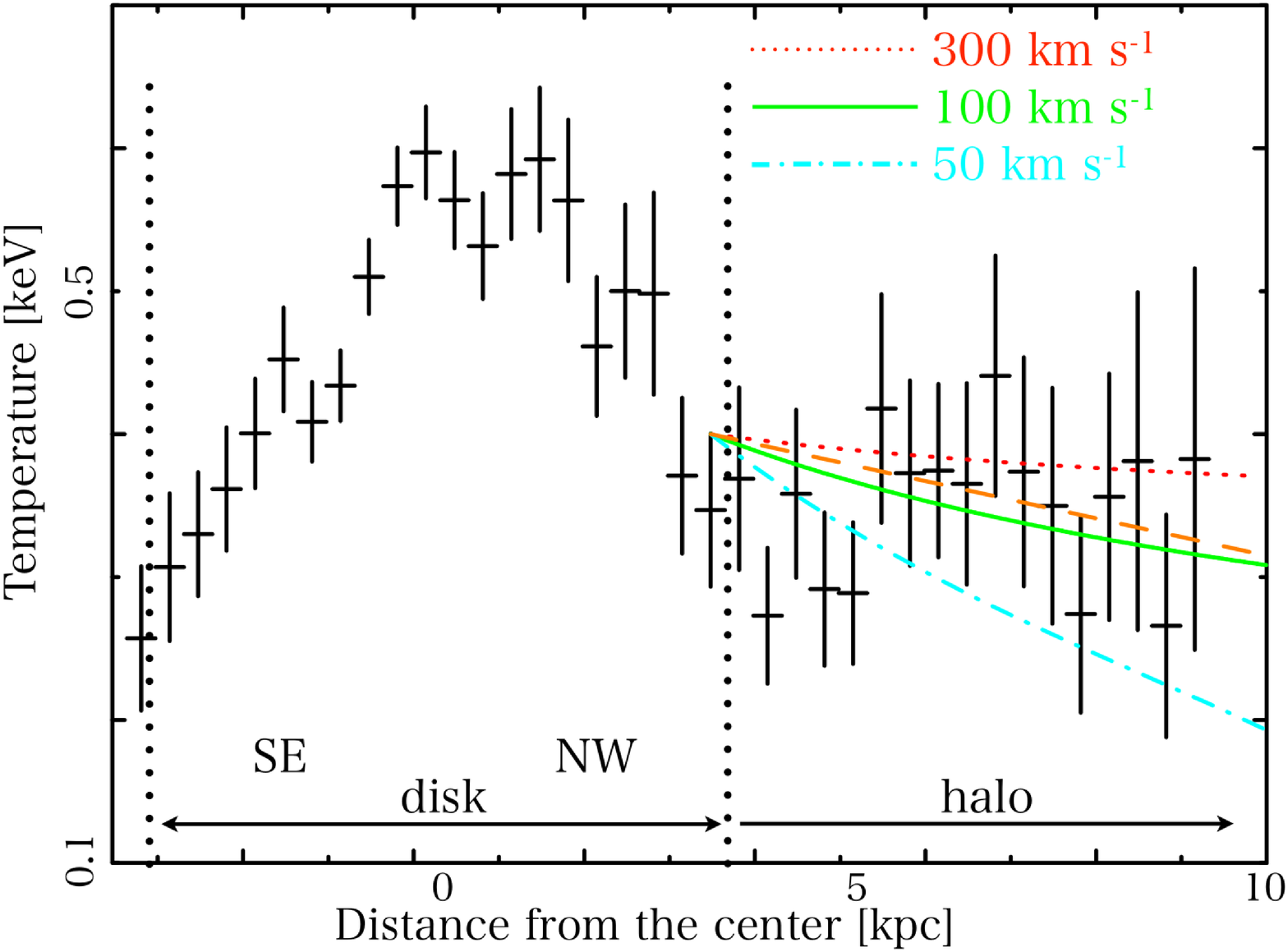} 
   \end{center}
   \caption{Comparison between the observed temperature profile and the expected temperature profile 
   in the halo of the {\it SB} region 
   through the radiative cooling process with constant velocities of 300 km s$^{-1}$ (dot), 100 km s$^{-1}$ (solid) 
   and 50 km s$^{-1}$ (dot-dash), respectively.
   Dashed line indicates the observed lowest temperature gradient within the 90 \% confidence level.}
\label{fig:velocity-constraint}
   \end{figure*}

In \S\ref{SEC:dynamics-of-the-hos-ism-gas}, 
it is suggested that the hot gas expands freely 
from the inner part of the halo towards the outer part of the halo as the outflow and 
no temperature gradient is detected in the halo of the {\it SB} region, 
which indicates that any effective cooling processes does not work significantly.
Even through the density in the halo is small, the X-ray emitting gas cools throughout the radiative cooling, 
which requires that the hot gas has to proceed at a certain level of the velocity towards the outer part of the halo  
to reproduce the observed flat temperature profile.
The temperature gradient in the halo of the {\it SB} region was calculated to be 
(0.3$\pm$1.6)$\times$10$^{-2}$ [keV / kpc] + 0.32$\pm$0.09 [keV].
To extract the lower limit of the velocity of hot interstellar gas, 
we adopted the slope of -1.3$\times$10$^{-2}$ [keV / kpc] as a 90 \% lower limit and 
0.40 keV as a starting point.
Assuming the following three conditions, 
we constrained the velocity of the hot interstellar gas in this section: 
(1) the ISM gas moves with a constant velocity in the halo, 
(2) the ISM gas cools through the radiative cooling process and 
(3) the density profile was estimated from the observed surface brightness profile in 0.4--0.8 keV.
In this model, during a specific time period $\Delta t$ [s], 
hot gas loses an energy of $\sim\Delta t\times n(R)\Lambda(T)$ 
leading to a temperature decrease satisfying the following relation: 
\begin{equation}
\label{eq:deltat}
T(t + \Delta t) = T - \Delta t\frac{n(R)\Lambda(T)}{k}, 
\end{equation}
where $k$, $n(R)$ and $\Lambda(T)$ are Boltzmann constant, the density of hot gas and 
cooling rate in units of erg~K$^{-1}$, cm$^{-3}$ and erg cm$^{-3}$ s$^{-1}$, respectively. 
A density profile is obtained from the surface brightness profile in 0.4--0.8 keV 
assuming a diameter of 5.2$'$; and a height of 21$"$ for a bin.
We adopt a cylindrical emitting region for a steady state outflow.
%
The density profile is exhibited 
in Figure \ref{fig:suzaku-xis1-04-08keV-density-profile}.
Resulting density in both the central region and the halo 
are consistent with previous works 
within a factor of 2 \citep{ngc253-nuclear-xmm,ngc253-xmm-rgs,ngc253-xmm-halo}.
We confirmed that 
systematic errors of the density due to the assumption of spectral parameters 
are less than 10 \%. 
The density profile was represented empirically with the double exponential model
as shown in Figure \ref{fig:suzaku-xis1-04-08keV-density-profile} upper right.
This smoothed empirical density function and radiative cooling rate by \citet{1993ApJS...88..253S} 
are used for the calculation of temperature decrease as show in equation (\ref{eq:deltat}).
We utilized cooling rate assuming CIE with the abundance ratio of [Fe/H] = -0.5 
considering our spectral analysis (see Table \ref{table:suzaku-xis0-1-3-spec_05-5keV_offset-halo_1t-2t}).
We interpolated emissivity between discrete values assuming linearity.
A comparison between the observed and the expected temperature profiles is shown 
in Figure \ref{fig:velocity-constraint}.
We also plotted the observed lowest temperature gradient within 90 \% confidence level.
Thus, it is suggested that the velocity of 100 km s$^{-1}$ is required as the lower limit 
to reproduce the observed temperature profile.
Considering the escape velocity of NGC 253, $\sim$220 km s$^{-1}$ \citep{ngc253-distance-2.58Mpc,ngc253-Halpha-obs}, 
it is suggested that once the ISM gas reaches the edge of the halo it can escape from the host galaxy 
breaking out its gravitational potential by adding the thermal velocity to the outflow velocity.
It should be noted that $\eta<1$ is naturally expected, although the exact value is hard to know 
in particular in the halo region.
Actually, the optical image of NGC 253 
\citep{2000MNRAS.314..511S,2009ApJ...701.1636M,2011MNRAS.414.3719W} suggests that 
the gas is clumpy at least in the disk region.
Recent numerical simulations (e.g., \citet{2000MNRAS.314..511S,2009ApJ...703..330C}) also indicates clumpy structure.
In the case of $\eta<1$, the required velocity is even larger 
because the density and hence the cooling rate increases with $\eta<1$.
\section{Conclusions}
\label{SEC:conclusions}
X-ray properties of the hot interstellar gas in starburst galaxy NGC 253 was investigated 
through abundance patterns, the surface brightness and the hardness ratio. 
We conducted spectral analysis for three regions in NGC 253, i.e., the $superwind$ region, the $disk$ region and the $halo$ region, 
characterized by multiwavelength observations.
Various emission lines such as O, Ne, Mg, Si and Fe were observed in all three regions.
We extracted abundance patterns of O/Fe, Ne/Fe, Mg/Fe and Si/Fe.
Resulting abundance patterns of all four elements are consistent in the three regions   
within the statistical error, which suggests that the origin of the hot interstellar gas in three regions 
is identical.
Abundance patterns in the $halo$ region are heavily contaminated by type II supernovae and   
therefore it is indicated that the hot interstellar gas in the $halo$ region is provided by the central starburst activity.
The energetics considering the observed luminosity and thermal energy in the $halo$ region also can support this indication 
on condition that 0.01--50 $\eta^{1/2}$ \% of the total emission in the nuclear region has flowed to the $halo$ region.

The hardness ratio and the surface brightness in the {\it SB} region 
show a different behavior between the disk and the halo.
The polytropic index ($\gamma$) of 5/3 in $T$$\rho^{1-\gamma}$ = const relation seems to be preferable 
at the hot disk of $<$3 kpc from the center along with the minor axis  
while no gradient was seen in the halo whose distance is between 3 and 10 kpc from the center  
but the surface brightness decreased gradually towards the outside.
These results indicate the following physical picture that the hot interstellar gas expands adiabatically 
in the disk and then moves as free expansion from the inner part of the halo towards the outer part of the halo as the outflow.
Finally, we constrained the velocity of this outflow in the halo. 
Assuming that the hot gas cools through radiative cooling and moves with the constant velocity 
in the halo, the velocity of $>$100 km s$^{-1}$ is required to reproduce the observed temperature profile.
Considering the escape velocity $\sim$220 km s$^{-1}$ of NGC 253, 
it is suggested that if the hot interstellar gas reaches the edge of the halo region 
hot interstellar gas can escape from the gravitational potential of NGC 253 as the outflow 
by combining the outflow velocity and the thermal velocity.
\section*{Acknowledgement}
I. M. is grateful to Kentaro Someya and Prof. Kazuhisa Mitsuda 
for useful advice of spectral analysis to extract abundance patterns and 
discussion to constrain the outflow velocity. 
Part of this work was financially supported by the Ministry of Education, Culture, Sports, Science and 
Technology, Grant-in Aid for Scientific Research 
10J07487, 15340088 and 22111513.

 
%


\end{document}